\documentclass[10pt]{iopart}
\pdfoutput=1
\usepackage{graphicx}
\usepackage{amssymb}
\eqnobysec

\newcommand{\beq}{\begin{equation}}
\newcommand{\eeq}{\end{equation}}
\newcommand{\be}{\begin{equation*}}
\newcommand{\ee}{\end{equation*}}
\newcommand{\beqa}{\begin{eqnarray}}
\newcommand{\eeqa}{\end{eqnarray}}
\newcommand{\bea}{\begin{eqnarray*}}
\newcommand{\eea}{\end{eqnarray*}}

\renewcommand{\th}{{\theta}}

\renewcommand{\l}{\ell}
\newcommand{\comport}[2]{\mathrel{\mathop{#1}\limits^{#2}}}
\newcommand{\stackunder}[2]{\mathrel{\mathop{#2}\limits_{#1}}}
\newcommand{\mean}[1]{\langle#1\rangle}
\newcommand{\prob}{\mathop{\rm Prob}\nolimits}
\newcommand{\dd}{{\rm d}}
\newcommand{\erf}{\mathop{\rm erf}}
\newcommand{\frad}[2]{\displaystyle{\displaystyle#1\over\displaystyle#2}}

\newcommand{\eps}{{\varepsilon}}
\newcommand{\w}{\widetilde}
\newcommand{\F}{{\cal F}}
\renewcommand{\L}{{\bf L}}
\renewcommand{\S}{{\bf S}}
\newcommand{\Y}{{\cal Y}}
\newcommand{\pe}{{(p)}}
\newcommand{\1}{{(1)}}
\newcommand{\2}{{(2)}}
\newcommand{\3}{{(3)}}
\newcommand{\4}{{(4)}}
\newcommand{\cy}{{\rm cyclic}}

\begin{document}

\title{Records for the moving average of a time series}
\author{Claude Godr\`eche and Jean-Marc Luck}
\address{
Institut de Physique Th\'eorique, Universit\'e Paris-Saclay, CEA and CNRS,
91191 Gif-sur-Yvette, France}\smallskip

\begin{abstract}
We investigate how the statistics of extremes and records is affected when
taking the moving average over a window of width $p$
of a sequence of independent, identically distributed random variables.
An asymptotic analysis of the general case,
corroborated by exact results for three distributions (exponential, uniform, power-law with unit exponent),
evidences a very robust dichotomy, irrespective of the window width,
between superexponential
and subexponential distributions.
For superexponential distributions the statistics of records is asymptotically unchanged by taking the moving average,
up to interesting distribution-dependent corrections to scaling.
For subexponential distributions
the probability of record breaking at late times is increased by a universal factor $R_p$,
depending only on the window width.
\end{abstract}

\eads{\mailto{claude.godreche@ipht.fr},\mailto{jean-marc.luck@ipht.fr}}

\section{Introduction}

When monitoring a time series, a feature which immediately attracts the attention of the observer is the sequence of
record values, viz., the successive largest or smallest values in the series \cite{chandler,renyi1,renyi2}.
The first example which comes to mind are weather records, i.e., the extreme occurrences of weather phenomena
such as the coldest or hottest days, the most rainy or windy days, and so on, for which studies abound
(see \cite{glick,redner,krug,coumou,coumou2,wergen} and references therein).
Other examples of records encountered in diverse complex physical systems are reviewed in \cite{revue},
to which the reader is referred for a comprehensive list of references.

The simplest situation to analyse is when the data are samples of a sequence
of independent, identically distributed (iid) random variables.
In such an instance
much is known on the statistics of records \cite{chandler,renyi1,renyi2,glick,arnold,nevzorov},
whose basics are easy to grasp.
Consider a sequence of iid continuous random variables $X_1,X_2, \dots$,
with common distribution function $F(x)=\prob(X<x)$ and density $f(x)={\rm d}F(x)/{\rm d}x$.
Throughout the following we assume that the $X_i$ are positive.
A record is said to occur at step $n$ if $X_n$ is larger than all previous variables, i.e., if
\be
X_n>L_{n-1}=\max(X_1,X_2,\dots,X_{n-1}),
\ee
where $L_n$ denotes the largest $X_i$ amongst the first $n$ random variables.
The probability of this event, or probability of record breaking,
\be
Q_n=\prob(X_n>L_{n-1}),
\ee
equals
\beq
Q_n=\frac{1}{n}
\label{qniid},
\eeq
as a consequence of the fact that the random variables $X_i$ are exchangeable \cite{renyi1,renyi2}.
The number $M_n$ of records up to time $n$ takes the values $1,\dots,n$ and can be expressed as the sum
\beq
M_n=I_1+I_2+\cdots+I_n,
\label{mind}
\eeq
where the indicator variable $I_n$ is equal to 1 if $X_n$ is a record and to 0 otherwise.
Taking the average, we have $\mean{I_i}=Q_i=1/i$, and so
\beq
\mean{M_n}=\sum_{i=1}^n \frac{1}{i}=H_n\approx\ln n+\gamma,
\label{Mniid}
\eeq
where $H_n$ is the $n$th harmonic number and $\gamma=0.577215\dots$ is Euler's constant.
It is a simple matter to show that the indicator variables $I_1,I_2,\dots,I_n$
are statistically independent \cite{renyi1,renyi2,revue}.
The distribution of $M_n$ ensues from this fact by elementary considerations
(see also section~\ref{sec:psubex} below).
The simple expression~(\ref{qniid}) of the probability of record breaking
and the full distribution of $M_n$ are universal,
in the sense that they do not depend on the underlying distribution $f(x)$.
From this standpoint the statistics of records for iid random variables exhibits a high degree of degeneracy.
In contrast, the statistics of the extreme value $L_n$ is distribution dependent, as is well known \cite{gnedenk}.

In the present work we investigate the statistics of records for sequences made of sums
of $p$ successive iid positive random variables, defined as follows.
For $p=2$,
\beq\label{eq:Y2}
Y_2=X_1+X_2,\quad Y_3=X_2+X_3,\dots, \quad Y_n=X_{n-1}+X_n,\ \dots,
\eeq
for $p=3$,
\beq\label{eq:Y3}
Y_3=X_1+X_2+X_3,\dots, \quad Y_n=X_{n-2}+X_{n-1}+X_n,\ \dots,
\eeq
or more generally,
\beq\label{eq:Yp}
Y_p=X_1+\cdots+X_p,\dots,\quad Y_n=X_{n-p+1}+\cdots+X_n, \ \dots
\eeq
The first terms of these sequences, which have not been written down explicitly,
may be omitted in the analysis of records.
For instance, in (\ref{eq:Y2}), $Y_1=X_1$ is always smaller than $Y_2$.
In (\ref{eq:Y3}), $Y_1=X_1$ and $Y_2=X_1+X_2$ are always smaller than $Y_3$, and so on.

Up to a normalisation, each of these sequences can be seen as
the \textit{moving average} of the sequence of iid variables $X_1,X_2,\dots$, defined as the mean of the last $p$ terms.
For instance the moving average with $p=2$ is
\be
\frac{X_1+X_2}{2}, \quad \frac{X_2+X_3}{2},\cdots,\quad \frac{X_{n-1}+X_n}{2},\ \dots
\ee
Taking the moving average is
a well-known method to analyse time series, which is equivalent to making a convolution of the signal by a square window, thus smoothing the signal.
For instance, instead of looking at the daily temperature at a given location, one can take the moving average over a period of one week, corresponding to choosing $p=7$.
The question posed here amounts therefore to knowing how records are affected by taking such an average.
The normalisation by the factor $p$ does not affect the outcome of the subsequent analysis.

Here the focus will be essentially on the particular case $p=2$.
Keeping the same notations as for the iid case, we shall primarily investigate the probability of record breaking,
\beq\label{eq:Qndef}
Q_n=\prob(Y_n>L_{n-1}),
\eeq
where $L_n$ denotes the largest $Y_i$ amongst the first $n$ ones,
\beq\label{eq:Ln}
L_n=\max(Y_1,Y_2,\dots,Y_n),
\eeq
and the mean number of records up to $n$,
\beq\label{eq:Mnaverage}
\mean{M_n}=\sum_{i=2}^{n}Q_i,
\eeq
where records are counted from the first complete sum $Y_2$ onwards.
As we shall see, these quantities are now sensitive to the choice of the underlying distribution $f(x)$ of the parent random variables $X_i$.
On the one hand, this does not come as a surprise since the new variables $Y_i$ are no longer exchangeable,
and the occurrences of records at various places are no longer independent.
On the other hand, it is yet slightly paradoxical that the degeneracy induced by the exchangeability
of the iid parent random variables~$X_i$ is now removed, revealing features of their common distribution,
since by taking the moving average one could have expected a loss of information instead.
We shall also study some features of the distribution of $L_n$.

\begin{figure}[!ht]
\begin{center}
\includegraphics[angle=0,width=.7\linewidth]{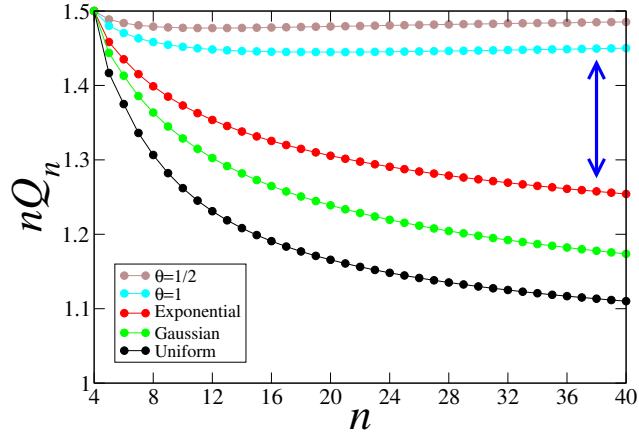}
\caption
{Plot of $n$ times the probability of record breaking $Q_n$ against~$n\ge4$
for several probability distributions $f(x)$.
From top to bottom: power-law distributions with tail index $\th=1/2$ and $\th=1$,
exponential, half-Gaussian, and uniform distributions (see table~\ref{tab:distribs}).}
\label{fig:nqn}
\end{center}
\end{figure}

In a nutshell, the main outcome of this work is as follows.
We find that the product~$nQ_n$ has only two possible limits for $p=2$,
depending on the class of distribution~$f(x)$, namely
\beq\label{eq:lim1}
nQ_n\to1
\eeq
for superexponential distributions,
that is, distributions either having a bounded support
or falling off faster than any exponential,
whereas
\beq\label{eq:lim32}
nQ_n\to\frac{3}{2}
\eeq
for subexponential distributions, whose tails decrease more slowly than any exponen\-tial.
The pure exponential distribution belongs to the first class, albeit marginally.
Figure~\ref{fig:nqn} shows a plot of $nQ_n$ against $n\ge4$
for all the examples of probability distributions $f(x)$
considered in the present paper (see table \ref{tab:distribs}).
Each dataset is the outcome of the numerical generation of $10^{10}$ sequences.
The vertical arrow underlines that the dichotomy between~(\ref{eq:lim1}) and~(\ref{eq:lim32})
becomes more and more visible as $n$ increases.
The values of $Q_n$ for $n=2,3,4$ are universal,
i.e., independent of the underlying distribution $f(x)$
(see section~\ref{sec:firstn}).

For higher values of the window width $p$,
denoting the probability of record breaking by $Q_n^{(p)}$,
(\ref{eq:lim1}) still holds for superexponential distributions, i.e.,
\beq\label{eq:lim1p}
nQ^{(p)}_n\to1,
\eeq
while, for subexponential distributions, (\ref{eq:lim32}) becomes
\beq\label{eq:lim32p}
nQ^{(p)}_n\to R_p,
\eeq
where the $R_p$ are universal rational numbers given by
\beq\label{eq:Rp1}
R_p=
\frac{3}{2},\ \frac{15}{8},\ \frac{35}{16},\ \frac{315}{128},\ \frac{693}{256},\dots
\eeq
for $p=2,3,4,5,6,\dots$,
and obtained by means of the Sparre Andersen theorem.

\begin{table}[!ht]
\begin{center}
\begin{tabular}{|l|c|c|c|}
\hline
Distribution & $F(x)$ & $f(x)$ & support\\
\hline
Uniform & $x$ & $1$ & $0<x<1$\\
\hline
Exponential & $1-\e^{-x}$ & $\e^{-x}$ & $x>0$\\
\hline
Half-Gaussian & $\erf x$ & $\frac{2}{\sqrt{\pi}}\,\e^{-x^2}$ & $x>0$\\
\hline
Power-law ($\th>0$) & $1-x^{-\th}$ & $\th\,x^{-1-\th}$ & $x>1$\\
\hline
\end{tabular}
\caption
{Distribution function $F(x)$, density $f(x)$ and support
of the examples of probability distributions considered in this work.}
\label{tab:distribs}
\end{center}
\end{table}

The setup of this paper is as follows.
Sections~\ref{sec:theo} to~\ref{sec:gal} concern the case $p=2$.
In section~\ref{sec:theo} we present the general setting
which will be used in all subsequent exact or asymptotic developments.
The next three sections
are devoted to exact analytical solutions of the problem for three distributions:
the exponential distribution (section~\ref{sec:exp}),
the uniform distribution (section~\ref{sec:uni}),
and the power-law distribution with index $\th=1$ (section~\ref{sec:th1}).
In order to compare the probability of record breaking to its universal value $Q_n=1/n$
in the iid situation (see~(\ref{qniid})),
we set
\beq\label{eq:deldef}
nQ_n=1+\delta_n.
\eeq
The exponential distribution appears as a marginal case where~(\ref{eq:lim1}) holds,
albeit with a logarithmic correction
\be
\delta_n\approx\frac{1}{\ln n}.
\ee
For the uniform distribution $\delta_n$ falls off as
\be
\delta_n\approx\sqrt\frac{\pi}{8n},
\ee
whereas (\ref{eq:lim32}) holds for the power-law distribution with $\th=1$.
A heuristic asymptotic analysis of the general case is then performed in section~\ref{sec:gal},
where the dichotomy between~(\ref{eq:lim1}) and~(\ref{eq:lim32}) is explained in simple terms,
and an estimate for the relative correction $\delta_n$ is derived.
Higher values of the window width $p$ are considered in section~\ref{sec:p} along the same line of thought.
The overall picture, including the dichotomy between~(\ref{eq:lim1}) and~(\ref{eq:lim32}), remains unchanged.
The non-trivial limit~$3/2$ in~(\ref{eq:lim32}) is replaced by the $p$-dependent
but otherwise universal limit $R_p$ (\ref{eq:Rp1}).
Section~\ref{sec:disc} contains a brief discussion of our findings.

Let us finally mention that the statistics of persistent events for the sequence~(\ref{eq:Y2})
has been studied in \cite{satya1,satya2,satya3}, for the case where the parent variables $X_i$ have a symmetric distribution $f(x)$.

\section{General setting}
\label{sec:theo}

This section sets the basis of all subsequent exact or asymptotic developments.
Hereafter and until the end of section~\ref{sec:gal}
we focus our attention on the sequence~(\ref{eq:Y2}) of sums of two terms.
Higher values of the width $p$ will be considered in section~\ref{sec:p}.

\subsection{Recursive structure}

We start by highlighting the recursive structure of the problem.
The first two maxima are necessarily $L_1=Y_1=X_1$ and $L_2=Y_2=X_1+X_2$.
The next maxima obey the recursion
\beq\label{eq:recursion}
L_n=\max(L_{n-1},Y_n)
=\left\{
\begin{array}{ll}
Y_n &\textrm{if } Y_n>L_{n-1},
\vspace{4pt}\\
L_{n-1} &\textrm{if } Y_n<L_{n-1}.
\end{array}
\right.
\eeq
This recursion should be understood as follows.
Starting from the couple of random variables $(L_{n-1},X_{n-1})$,
one draws the random variable $X_n$,
which is independent of~$L_{n-1}$ and $X_{n-1}$,
and sets $Y_n=X_{n-1}+X_n$.
This generates the new $L_n$, or alternatively the new couple $(L_n,X_n)$:
\be
(L_{n-1},X_{n-1})\comport{\leadsto}{X_n} (L_{n},X_{n}).
\ee
In other words, at each time step $n$, the newly drawn random variable $X_n$
acts as a noise on the dynamics of the couple $(L_{n-1},X_{n-1})$.
The value of $L_n$ depends on the branch of the recursion, denoted
respectively by $\L$ (for larger) and $\S$ (for smaller):
\beqa\label{eq:branches}
(\L): Y_n>L_{n-1}\Longrightarrow L_n=Y_n,
\nonumber\\
(\S): Y_n<L_{n-1}\Longrightarrow L_n=L_{n-1}.
\eeqa
In the first case, $Y_n$ is a record since it satisfies
\be
Y_n>\max(Y_1,\dots, Y_{n-1}).
\ee
This event occurs with probability $Q_n$ (see~(\ref{eq:Qndef})).
Hereafter we make use of the recursion~(\ref{eq:recursion})
to derive the key relations~(\ref{eq:recurF}) and~(\ref{eq:start})
for the functions $F_n(\l,x)$ introduced in~(\ref{eq:fndef}).

Let us mention that a similar,
albeit simpler recursive scheme applies to the theory of records for iid random variables.

\subsection{Basic quantities}
\label{sec:observables}

Starting from the joint distribution function of the couple of random variables
$(L_n,X_n)$,
\be
\prob(L_n<\l,X_n<x),
\ee
and taking its derivative with respect to $x$, yields the quantity
\be
F_n(\l,x)=\partial_{x}\prob(L_n<\l,X_n<x),
\ee
which plays a central role in the present work.
It is equivalently defined as
\beq\label{eq:fndef}
F_n(\l,x){\rm d}x=\prob(L_n<\l,x<X_n<x+{\rm d}x).
\eeq
The underlying distribution $f(x)$ is recovered in the $\l\to\infty$ limit:
\be
F_n(\infty,x)=f(x).
\ee
By differentiating $F_n(\l,x)$ with respect to $\l$, one gets the joint
probability density of the couple $(L_{n},X_{n})$:
\be
f_n(\l,x)=\partial_{\l} F_n(\l,x),\quad F_n(\l,x)=\int_0^\l{\rm d}\l'\,f_n(\l',x).
\ee
Conversely, integrating on the second variable restores
\be
\prob(L_n<\l,X_n<x)=\int_{0}^{x}{\rm d}x'\,F_n(\l,x').
\ee
In particular the distribution function of the maximum $L_n$ is obtained when the integral
runs over its full range (i.e., $x=\l$):
\beq\label{FLint}
\F_n(\l)=\prob(L_n<\l)=\int_{0}^\l{\rm d}x'\,F_n(\l,x').
\eeq
Its derivative with respect to $\l$ yields the density $f_{L_n}(\l)$.
The determination of the mean maximum ensues:
\beq\label{eq:Lave}
\mean{L_n}=\int_0^\infty\dd\l(1-\F_n(\l)).
\eeq
Finally, the normalization of the joint density $f_n(\l,x)$ implies
\beq\label{eq:normfnlx}
\int_0^\infty\dd\l\int_0^\l{\rm d}x\,f_n(\l,x)=\int_0^\infty{\rm d}x\int_x^\infty\dd
\l\,f_n(\l,x)=1.
\eeq

\subsection{First values of $n$}

The quantities defined above have explicit expressions for $n=1$ and 2 in full generality.

For $n=1$ we have
\be
\prob(L_1<\l,X_1<x)=\prob(X_1<x)=F(x),
\ee
whenever $x<\l$, since $L_1=X_1$.
Differentiating with respect to $x$ gives
\beq\label{eq:F1}
F_1(\l,x)=f(x)
\eeq
and
\be
\F_1(\l)=F(\l).
\ee

For $n=2$,
knowing that $L_2=Y_2=X_1+X_2$ allows one to compute
\be
\prob(L_2<\l,X_2<x)=F(x)F(\l-x)
+\int_{\l-x}^\l{\rm d}x_1\, f(x_1)F(\l-x_1),
\ee
from which $F_2(\l,x)$ ensues by derivation with respect to $x$:
\beq\label{eq:F2}
F_2(\l,x)=f(x)F(\l-x),
\eeq
consistently with the definition (with informal notation)
\be
\prob(L_2=X_1+X_2<\l,X_2=x)=\prob(X_1<\l-x,X_2=x).
\ee
Then, taking a derivative with respect to $\l$, we have
\be
f_2(\l,x)=f(x)f(\l-x),
\ee
and finally
\be
\F_2(\l)=\int_{0}^\l{\rm d}x\,f(x)F(\l-x).
\ee

\subsection{Recursion relation for the function $F_n(\l,x)$}

\begin{figure}[!ht]
\begin{center}
\includegraphics[angle=0,width=1.\linewidth]{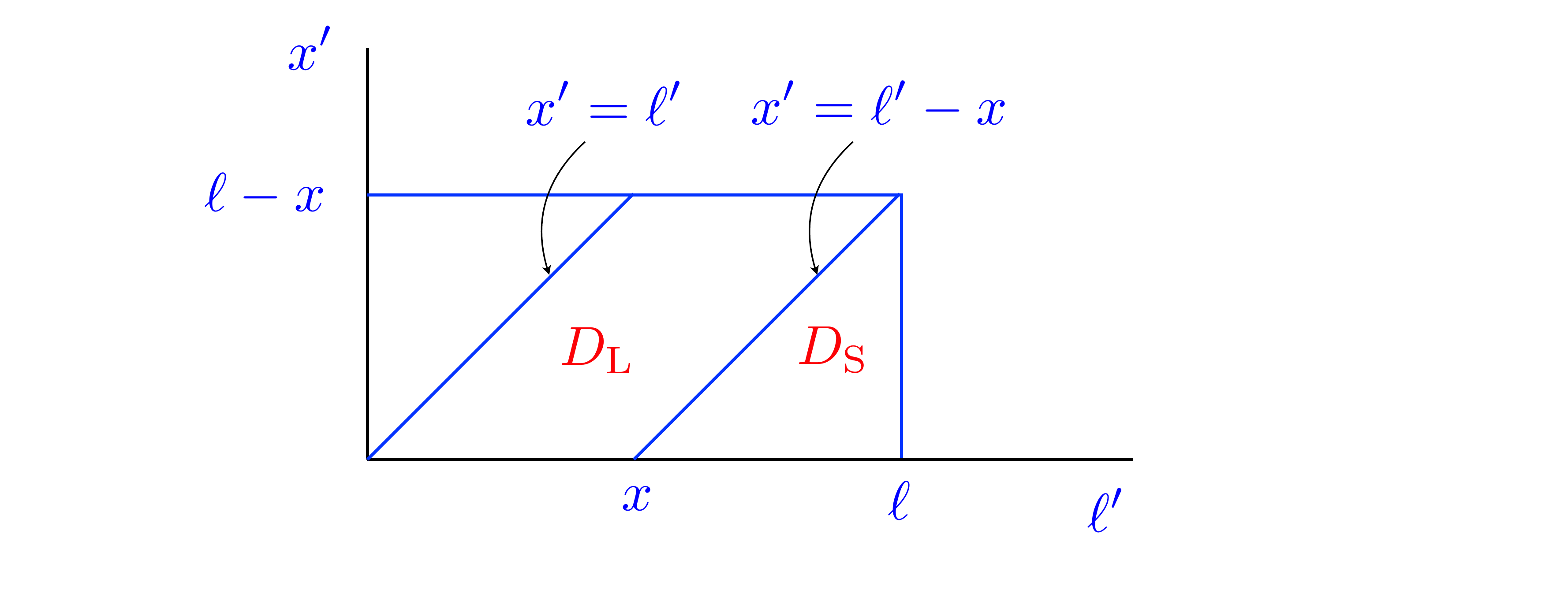}
\caption
{The domains $D_\L$ and $D_\S$ in the $(\l',x')$ plane.}
\label{fig:domaines}
\end{center}
\end{figure}

The recursion~(\ref{eq:recursion}) implies
\beq\label{eq:recurFn}
F_n(\l,x)=
f(x)\int\dd\l'{\rm d}x'\,f_{n-1}(\l',x')\Theta(\l-\max(\l',x'+x)),
\eeq
where $\Theta$ denotes Heaviside function.
The right-hand side of this equation decomposes into two contributions, associated
to the two branches $\L$ and $\S$,
\beq\label{eq:FnlxLS}
\fl F_n(\l,x)
=f(x)\int_{D_\L}\dd\l'{\rm d}x'\,f_{n-1}(\l',x')
+f(x)\int_{D_\S}\dd\l'{\rm d}x'\,f_{n-1}(\l',x'),
\eeq
where the domains $D_\L$ and $D_\S$, depicted in figure~\ref{fig:domaines}, are
respectively defined as
\bea
D_\L=\{\l'<x+x'<\l\},
\\
D_\S=\{x+x'<\l'<\l\},
\eea
hence
\beqa\label{eq:FDL}
\int_{D_\L}\dd\l'{\rm d}x'\,f_{n-1}(\l',x')=\int_{0}^{\l-x}{\rm d}x'\int_{x'}^{x+x'}\dd
\l'\,f_{n-1}(\l',x'),
\\
\int_{D_\S}\dd\l'{\rm d}x'\,f_{n-1}(\l',x')=\int_{0}^{\l-x}\dd
x'\int_{x+x'}^{\l}\dd\l'\,f_{n-1}(\l',x').
\label{eq:FDS}
\eeqa

Adding these two contributions yields
\be
F_n(\l,x)=f(x)\int_{0}^{\l-x}{\rm d}x'\int_{x'}^{\l}\dd\l'\,f_{n-1}(\l',x'),
\ee
which vanishes whenever its two arguments are equal ($n\ge2$):
\beq\label{eq:Fll}
F_n(\l,\l)=0.
\eeq
We thus obtain the following recursion relation for the function $F_n(\l,x)$:
\beq\label{eq:recurF}
F_n(\l,x)=f(x)\int_{0}^{\l-x}{\rm d}x'\,F_{n-1}(\l,x').
\eeq
This equation and its differential form
\beq\label{eq:start}
F_n'(\l,x)=\frac{f'(x)}{f(x)}F_n(\l,x)-f(x)F_{n-1}(\l,\l-x),
\eeq
obtained by differentiating~(\ref{eq:recurF}) with respect to $x$
\footnote{Throughout the following, accents on functions denote their (partial)
derivatives with respect to $x$.},
are key formulas of this work
and the starting points of many subsequent developments.

As a consequence of~(\ref{eq:recurF}), we have ($n\ge2$)
\beq\label{eq:Fzero}
F_n(\l,0)=f(0)\F_{n-1}(\l).
\eeq
Finally, differentiating~(\ref{eq:FnlxLS}) with respect to $\l$ yields
\beq\label{eq:recurf}
f_n(\l,x)=\L f_{n-1}(\l,x)+\S f_{n-1}(\l,x),
\eeq
with the notations
\beqa\label{eq:LetL}
\L f_{n-1}(\l,x)&=&f(x)\int_{\l-x}^{\l}\dd\l'\, f_{n-1}(\l',\l-x)
\nonumber\\
&=&f(x)F_{n-1}(\l,\l-x),
\\
\S f_{n-1}(\l,x)&=&f(x)\int_{0}^{\l-x}{\rm d}x'\,f_{n-1}(\l,x'),
\label{eq:LetS}
\eeqa
using~(\ref{eq:FDL}) and~(\ref{eq:FDS}).
Alternatively, differentiating~(\ref{eq:recurF}) with respect to $\l$ gives
\beq\label{eq:recurf2}
f_n(\l,x)=f(x)F_{n-1}(\l,\l-x)+f(x)\int_0^{\l-x}{\rm d}x'\,f_{n-1}(\l,x'),
\eeq
which is identical to~(\ref{eq:recurf}).

\subsection{Probability of record breaking}

The probability of record breaking $Q_n$ is the probability that the last
variable is larger than all previous ones (see~(\ref{eq:Qndef})),
\be
Q_n=\prob(Y_n>L_{n-1}).
\ee
This probability thus equals the weight of branch $\L$.
For, recalling~(\ref{eq:normfnlx}) and~(\ref{eq:recurf}),
\bea
1&=&\int_{0}^\infty\dd\l\,\int_{0}^{\l}{\rm d}x\, f_{n}(\l,x)
\\
&=&\int_{0}^\infty\dd\l\,\int_{0}^{\l}{\rm d}x\,\left(\L f_{n-1}(\l,x)+\S
f_{n-1}(\l,x)\right),
\eea
where the two terms corresponding respectively to the weights of the two
branches $\L$ and $\S$ are $Q_n$ and $1-Q_n$.
So the expression of $Q_n$ is ($n\ge2$)
\beqa\label{eq:Qn}
Q_n&=&\int_{0}^\infty\dd\l\,\int_{0}^{\l}{\rm d}x\,\L f_{n-1}(\l,x)
\nonumber\\
&=&\int_{0}^{\infty}\dd\l\int_{0}^{\l}{\rm d}x\, f(x)F_{n-1}(\l,\l-x)
\nonumber\\
&=&\int_{0}^{\infty}\dd\l\int_{0}^{\l}{\rm d}x\, f(\l-x)F_{n-1}(\l,x).
\eeqa

\subsection{Universal values of the probability of record breaking}
\label{sec:firstn}

The first few values of $Q_n$ are universal,
i.e., independent of the underlying distri\-bution~$f(x)$.
For $n=2$,
\beq\label{eq:Q2}
Q_2=1,
\eeq
since $Y_2=X_1+X_2$ is always larger that $Y_1=X_1$.
This result can be recovered by inserting~(\ref{eq:F1}) into~(\ref{eq:Qn}).
For $n=3$,
\beq\label{eq:Q3}
Q_3=\frac{1}{2},
\eeq
since $Y_3>Y_2$ is equivalent to $X_3>X_1$, which holds with probability $1/2$.
This result can be recovered by inserting~(\ref{eq:F2}) into~(\ref{eq:Qn}).
It turns out that for $n=4$, $Q_n$ has also a universal value,
\beq\label{eq:Q4}
Q_4=\frac{3}{8},
\eeq
irrespective of the distribution $f(x)$.
This can be demonstrated by a simple application of the Sparre Andersen theorem \cite{sparre1,sparre2,feller}.
This theorem states in particular that, for a sequence of iid variables $Z_n$ with a continuous symmetric distribution,
the probability that the first $n$ partial sums are all positive,
\be
P_n=\prob(Z_1>0,Z_1+Z_2>0,\dots,Z_1+Z_2+\cdots+Z_n>0),
\ee
is a universal rational number,
\beq\label{eq:psa}
P_n=\frac{1}{2^{2n}}{2n\choose n}
=\frac{(2n)!}{(2^nn!)^2}
=1,\ \frac{1}{2},\ \frac{3}{8},\ \frac{5}{16},\ \frac{35}{128},\ \frac{63}{256},\dots,
\eeq
for $n=0,1,2,3,4,5,\dots$,
with generating function
\beq\label{eq:piz}
\w P(z)=\sum_{n\ge0}z^nP_n=\frac{1}{\sqrt{1-z}}.
\eeq
In the present case, by definition,
\bea
Q_4&=&\prob(Y_4>Y_3,Y_4>Y_2)
\\
&=&\prob(X_4-X_2>0,X_4-X_2+X_3-X_1>0)
\\
&=&\prob(Z_1>0,Z_1+Z_2>0),
\eea
where the random variables $Z_1=X_4-X_2$ and $Z_2=X_3-X_1$
are iid and have a continuous symmetric distribution.
Therefore the theorem applies and $Q_4=P_2$, which is the result announced in (\ref{eq:Q4}).
It would be cumbersome to recover this directly by means of~(\ref{eq:Qn}).

The probability of record breaking $Q_n$ is no longer universal for $n\ge5$.
It is indeed clear from figure~\ref{fig:nqn}
that already $Q_5$ depends on the underlying distribution~$f(x)$.

It results from the foregoing that the first values of the mean number of records
(see~(\ref{eq:Mnaverage}))
\be
\mean{M_2}=1,\quad
\mean{M_3}=\frac{3}{2},\quad
\mean{M_4}=\frac{15}{8},
\ee
are also universal.

In the forthcoming sections we apply the general formalism presented in this section
to derive analytical solutions of the differential recursion~(\ref{eq:start})
for the exponential distribution (section~\ref{sec:exp}),
the uniform distribution (section~\ref{sec:uni}),
and the power-law distribution with index $\th=1$ (section~\ref{sec:th1}).

\section{Exponential distribution}
\label{sec:exp}

This section presents an exact solution of the problem
for the case of exponentially distributed random variables $X_i$,
with common density $f(x)=\e^{-x}$ and distribution function $F(x)=1-\e^{-x}$ (see table~\ref{tab:distribs}).

\subsection{Differential equations}

The exact solutions derived in this section and in the two subsequent ones
rely on the differential equation~(\ref{eq:start}), which reads, in the present case,
\beq\label{eq:FFprime}
F'_n(\l,x)+F_n(\l,x)=-\e^{-x}F_{n-1}(\l,\l-x).
\eeq
Differentiating once more yields ($n\ge3$)
\beq\label{eq:equadif}
F''_n(\l,x)+F'_n(\l,x)+\e^{-\l}F_{n-2}(\l,x)=0.
\eeq
This is a recursive differential equation in the variable $x$, while $\l$ plays the role of a parameter.
Setting $x=0$ in~(\ref{eq:FFprime}) gives ($n\ge3$)
\beq\label{eq:FFprime0}
F_n(\l,0)+F'_n(\l,0)=0,
\eeq
where the interpretation of the first term is given in (\ref{eq:Fzero}).

\subsection{First values of $n$}

For the first few values of $n$, we obtain
\be
F_1(\l,x)=\e^{-x},\quad F_2(\l,x)=\e^{-x}-\e^{-\l},
\ee
\be
F_3(\l,x)=\e^{-x}-(\l-x)\e^{-x-\l}-\e^{-\l}.
\ee
Inserting these expressions into~(\ref{eq:Qn}), we recover the universal results
for $Q_2$, $Q_3$ and $Q_4$ derived in section~\ref{sec:firstn}.
Equations~(\ref{eq:Fll}) and~(\ref{eq:FFprime0}) are complemented by
\beqa\label{eq:FFin}
F_1(\l,\l)=\e^{-\l}, \quad
F_1(\l,0)+F'_1(\l,0)=0,
\nonumber\\
F_2(\l,0)+F'_2(\l,0)=-\e^{-\l}.
\eeqa
We have also
\be
\F_1(\l)=1-\e^{-\l},\quad\F_2(\l)=1-(\l+1)\e^{-\l},
\ee
\be
\F_3(\l)=1-\e^{-2\l}-2\l\e^{-\l}\l.
\ee
Inserting these expressions into~(\ref{eq:Lave}) yields
$\mean{L_1}=1$,
$\mean{L_2}=2$ and
$\mean{L_3}=5/2$.

\subsection{Generating function}

In order to solve the recursive differential equation~(\ref{eq:equadif}) for all values of $n$,
we introduce the generating function
\beq\label{eq:Fgenedef}
\w F(z,\l,x)=\sum_{n\ge1}z^nF_{n}(\l,x),
\eeq
which satisfies (using~(\ref{eq:equadif}))
\be
\w F''(z,\l,x)+\w F'(z,\l,x)+z^2\e^{-\l}\w F(z,\l,x)=0,
\ee
the solution of which is
\beq\label{fgeneres}
\w F(z,\l,x)=A_+\,\e^{a_+x} +A_-\,\e^{a_-x},
\eeq
with
\be
a_{\pm}=\frac{1\pm w}{2},\quad w=\sqrt{1-4z^2\e^{-\l}}.
\ee
The amplitudes $A_{\pm}$ are determined by the boundary conditions
(see~(\ref{eq:FFin}))
\be
\w F(z,\l,0)+\w F'(z,\l,0)=-z^2\e^{-\l},\quad
\w F(z,\l,\l)=z\e^{-\l},
\ee
yielding
\be
A_{\pm}=
\pm z\,\frac{a_\pm\e^{-\l/2}+z\e^{\pm w\l/2-\l}}{w\cosh\frad{w\l}{2}-\sinh\frad{w\l}{2}}.
\ee

\subsection{Probability of record breaking}

Using~(\ref{eq:Qn}),
the generating function of the $Q_n$ reads
\beq\label{Qzexp}
\w Q(z)=\sum_{n\ge2}z^{n}Q_n=z\int_0^{\infty}\dd\l\,\e^{-\l}I(z,\l),
\eeq
with
\beqa\label{Izl}
I(z,\l)
&=&\int_{0}^{\l}{\rm d}x\,\e^{x}\w F(z,\l,x)
\nonumber\\
&=&A_{+}\frac{\e^{(1-a_{+})\l}-1}{1-a_{+}}
+A_{-}\frac{\e^{(1-a_{-})\l}-1}{1-a_{-}}
=\frac{N(z,\l)}{D(z,\l)},
\eeqa
and
\bea
\fl N(z,\l)\!=\!4z(1-z)\e^{-\l/2}+(1-w^2-4z)\cosh\frac{w\l}{2}
+(2(1+w^2)z+1-w^2)\frac{1}{w}\sinh\frac{w\l}{2},
\\
\fl D(z,\l)\!=\!(w^2-1)\left(\cosh\frac{w\l}{2}-\frac{1}{w}\sinh\frac{w\l}{2}\right).
\eea

The integral over $\l$ in~(\ref{Qzexp}) cannot be carried out in closed form.
By expanding~$I(z,\l)$ as a power series in $z$ and integrating term by term with respect to~$\l$,
we obtain the values of the probability of record breaking $Q_n$ and mean number of records $\mean{M_n}$
given in table~\ref{tab:exp} up to $n=8$.

\begin{table}[!ht]
\begin{center}
\begin{tabular}{|c|c|c|c||c|c|c|c|}
\hline
$n$ & 2 & 3 & 4 & 5 & 6 & 7 & 8\\
\hline
$Q_n$ & 1 & $\frad{1^{\vphantom{M}}}{2_{\vphantom{M}}}$ &
$\frad{3}{8}$ & $\frad{7}{24}$ & $\frad{155}{648}$ & $\frad{131}{648}$ & $\frad{14503}{82944}$\\
\hline
$\mean{M_n}$ & 1 & $\frad{3^{\vphantom{M}}}{2_{\vphantom{M}}}$ &
$\frad{15}{8}$ & $\frad{13}{6}$ & $\frad{1559}{648}$ & $\frad{845}{324}$ & $\frad{8549}{3072}$\\
\hline
\end{tabular}
\caption
{Exact values of the probability of record breaking $Q_n$ and mean number of records $\mean{M_n}$
up to $n=8$, for an exponential distribution.
Expressions to the left of the double bar are universal.}
\label{tab:exp}
\end{center}
\end{table}

The asymptotic decay of $Q_n$ at large $n$
can be derived as follows.
Setting $z=\e^{-\eps}$,~(\ref{Izl}) becomes
\be
I(z,\l)\approx\frac{\l-1}{\eps+(\l-1)\e^{-\l}},
\ee
in the relevant regime where $\eps$ and $\e^{-\l}$ are simultaneously small.
Inserting this into~(\ref{Qzexp}), and dealing with $n$ as a continuous variable,
we obtain the estimate
\be
\w Q(z)\approx\int_0^\infty{\rm d}n\,\e^{-n\eps}Q_n
\approx\int_1^\infty\dd\l\,\frac{(\l-1)\e^{-\l}}{\eps+(\l-1)\e^{-\l}}.
\ee
Performing the inverse Laplace transform yields
\beq\label{Qnint}
Q_n\approx\int_1^\infty\dd\l\,(\l-1)\,\exp\left(-\l-n(\l-1)\e^{-\l}\right).
\eeq
Setting
\beq\label{lamdef}
\lambda=\ln n
\eeq
and changing integration variable from $\l$ to $\mu$
such that $(\l-1)\e^{-\l}=\e^{-\mu}$, we obtain formally
\beq\label{nQnint}
nQ_n\approx
\int_{-\infty}^\infty\dd\mu\,\underbrace{\exp\left(\lambda-\mu-\e^{\lambda-\mu}\right)}
\left(1+\frac{1}{\l(\mu)-2}\right).
\eeq
The expression underlined by the brace is
the normalized Gumbel distribution with parameter $\lambda$.
This distribution is peaked around $\mu=\lambda$.
More precisely, considering the following average with respect to this distribution,
\be
\int_{-\infty}^\infty\dd\mu\,\e^{s\mu}\exp\left(\lambda-\mu-\e^{\lambda-\mu}\right)
=\e^{s\lambda}\Gamma(1-s),
\ee
we obtain
\be
\fl\int_{-\infty}^\infty\dd\mu\,\phi(\mu)\exp\left(\lambda-\mu-\e^{\lambda-\mu}\right)
=\phi(\lambda)+\gamma\phi'(\lambda)
+\left(\frac{\gamma^2}{2}+\frac{\pi^2}{12}\right)\phi''(\lambda)+\cdots,
\ee
for any slowly varying function $\phi(\mu)$,
where $\gamma$ is Euler's constant.
Applying this to the function inside the large parentheses in~(\ref{nQnint}),
we obtain the expansion
\beq\label{eq:lexpasy}
nQ_n=1+\delta_n=1+\frac{1}{\lambda}-\frac{\nu-2}{\lambda^2}
+\frac{\nu^2-5\nu+5+\pi^2/6}{\lambda^3}+\cdots,
\eeq
with the notation~(\ref{lamdef}) and
\be
\nu=\ln\lambda+\gamma=\ln\ln n+\gamma.
\ee
Omitting details,
let us mention that a similar analysis yields the following expansion
for the mean number of records up to time $n$:
\beq\label{eq:mexpasy}
\mean{M_n}=\lambda+\nu-1+\frac{\nu-1}{\lambda}
-\frac{\nu^2-4\nu+3+\pi^2/6}{2\lambda^2}+\cdots
\eeq
Equations (\ref{eq:lexpasy}) and (\ref{eq:mexpasy}) give the first few terms of asymptotic expansions
to all orders in $1/\lambda$.
The ambiguity in the formal expression~(\ref{nQnint})
originating in the pole at $\l=2$, i.e., $\mu=1$,
is indeed exponentially small in $\lambda$.

\subsection{Mean value of the maximum}

Using~(\ref{eq:Lave}),~(\ref{eq:Fzero}) and~(\ref{eq:Fgenedef}),
we obtain the generating function of the mean value $\mean{L_n}$
of the largest variable $Y_i$ up to time $n$,
\be
L(z)=\sum_{n\ge1}z^n\mean{L_n}
=\int_0^\infty\dd\l\,\left(\frac{1}{1-z}-\frac{1}{z}\w F(z,\l,0)\right).
\ee
The explicit expression~(\ref{fgeneres}) of the generating function $\w F(z,\l,x)$ implies
\be
\frac{1}{1-z}-\frac{1}{z}\w F(z,\l,0)=\frac{z\,\e^{-\l}}{1-z}\left(I(z,\l)+1\right),
\ee
so that
\be
L(z)=\frac{z+\w Q(z)}{1-z},
\ee
and finally
\beq\label{eq:Lnexp}
\mean{L_n}=\mean{M_n}+1.
\eeq
This remarkable identity between mean values is a peculiarity of the exponential distribution.
The first few values of $\mean{L_n}$ can therefore be read from table~\ref{tab:exp},
whereas its asymptotic growth can be read from~(\ref{eq:mexpasy}).
Let us notice that a similar identity, i.e., $\mean{L_n}=\mean{M_n}=H_n$,
holds for extremes and records of exponentially distributed iid random variables
(see \cite{bertin1,bertin2} for a discussion of related matters).

\section{Uniform distribution}
\label{sec:uni}

The case where the random variables $X_i$ are uniformly distributed on [0, 1],
with common density $f(x)=1$ and distribution function $F(x)=x$ for $0<x<1$ (see table~\ref{tab:distribs}),
also lends itself to an exact solution of the problem.

\subsection{Sectors}

\begin{figure}[!ht]
\begin{center}
\includegraphics[angle=0,width=1.\linewidth]{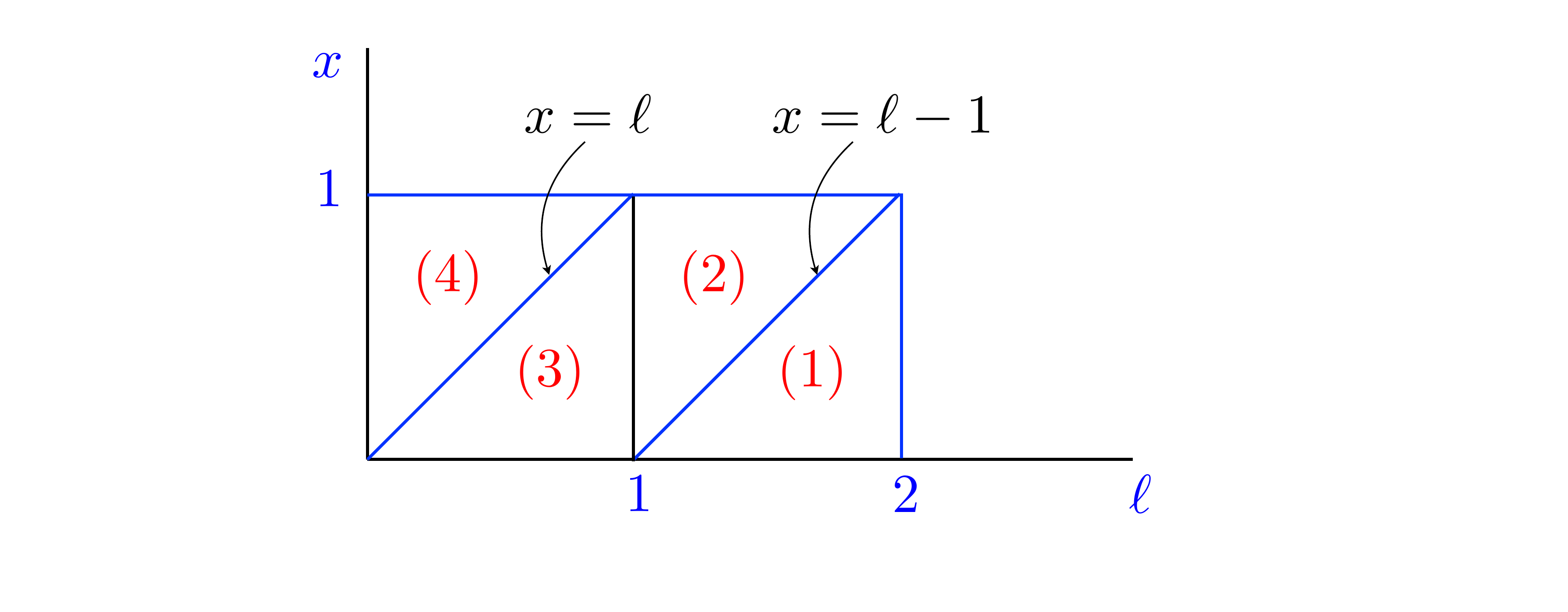}
\caption
{Four sectors in the $(\l,x)$ plane for the case of a uniform distribu\-tion~$f(x)$.}
\label{fig:fig-unif}
\end{center}
\end{figure}

Here, the relevant part of the $(\l,x)$ plane is the rectangle defined by $0<\l<2$ and $0<x<1$.
This region splits into four sectors (see figure~\ref{fig:fig-unif}):
\be
\left\{\matrix{
\1:\quad & 1<\l<2,\ \ 0<x<\l-1,\hfill\cr
\2:\hfill & 1<\l<2,\ \ \l-1<x<1,\hfill\cr
\3:\hfill & 0<\l<1,\ \ 0<x<\l,\hfill\cr
\4:\hfill & 0<\l<1,\ \ \l<x<1.\hfill
}\right.
\ee

The functions $F_n(\l,x)$ assume a priori different analytical forms in these four sectors.
The recursion~(\ref{eq:recurF}) reads
\beqa\label{eq:unifF123}
F_n^{\1}(\l,x)=\int_0^{\l-1}{\rm d}x' F_{n-1}^{\1}(\l,x')+\int_{\l-1}^{1}{\rm d}x'
F_{n-1}^{\2}(\l,x'),
\nonumber\\
F_n^{\2}(\l,x)=\int_0^{\l-1}{\rm d}x' F_{n-1}^{\1}(\l,x')+\int_{\l-1}^{\l-x}{\rm d}x'
F_{n-1}^{\2}(\l,x'),
\nonumber\\
F_n^{\3}(\l,x)=\int_0^{\l-x}{\rm d}x' F_{n-1}^{\3}(\l,x'),
\nonumber\\
F^{\4}_n(\l,x)=0.
\eeqa

The first function $F_n^{\1}(\l,x)$ is independent of $x$,
whereas the last one vanishes,
so the information of interest is contained in sectors~\2 and~\3.

The probability of record breaking reads
\be
Q_n=Q^{\2}_n+Q^{\3}_n,
\ee
with
\bea
Q^\2_n&=&\int_1^2 \dd\l\,\int_{\l-1}^1{\rm d}x\,F_{n-1}^{\2}(\l,x),
\\
Q^\3_n&=&\int_0^1 \dd\l\,\int_0^\l{\rm d}x\,F_{n-1}^{\3}(\l,x).
\eea
Similarly, the mean value of the maximum reads
\be
\mean{L_n}=2-I_n^\2-I_n^\3,
\ee
with
\bea
I^\2_n&=&\int_1^2 \dd\l\,F_{n+1}^{\2}(\l,\l-1),
\\
I^\3_n&=&\int_0^1 \dd\l\,F_{n+1}^{\3}(\l,0).
\eea

Finally, differentiating~(\ref{eq:unifF123}) with respect to $x$,
we obtain the following differential equations,
valid in both sectors $\2$ and $\3$:
\beqa\label{eq:derivUnif1}
F_n'(\l,x)=-F_{n-1}(\l,\l-x),
\\
F_n''(\l,x)=-F_{n-2}(\l,x).
\label{eq:derivUnif2}
\eeqa
These equations, which can alternatively be read off from~(\ref{eq:start}),
will be instrumental hereafter.
We thus obtain the following expressions for the first values of $n$:
\bea
F_1^{\1}(\l,x)=F_1^{\2}(\l,x)=F_1^{\3}(\l,x)=1
\eea
for $n=1$,
\bea
F_2^{\1}(\l,x)=1,
\\
F_2^{\2}(\l,x)= F_2^{\3}(\l,x)=\l-x
\eea
for $n=2$,
and
\bea
F_3^{\1}(\l,x)=\frac{1}{2}(-2+4\l-\l^2),
\\
F_3^{\2}(\l,x)=\frac{1}{2}(2\l-1-x^2),
\\
F_3^{\3}(\l,x)=\frac{1}{2}(\l^2-x^2)
\eea
for $n=3$.

\subsection{Analysis of sector~\3}

The generating function
\be
\w
F^{\3}(z,\l,x)=\sum_{n\ge1}z^nF^{\3}_n(\l,x)
\ee
satisfies
\be
\w F'^{\3}(z,\l,x)=-z\w F^{\3}(z,\l,\l-x),
\ee
because of~(\ref{eq:derivUnif1}),
\be
\w F''^{\3}(z,\l,x)=-z^2\w F^{\3}(z,\l,x),
\ee
because of~(\ref{eq:derivUnif2}), and
\beq\label{eq:condlim}
\w F^{\3}(z,\l,\l)=z,\quad\w F'^{\3}(z,\l,0)=-z^2,
\eeq
because of~(\ref{eq:Fll}).
Hence
\be
\w F^{\3}(z,\l,x)=A\cos zx+B\sin zx,
\ee
where the amplitudes $A$ and $B$, which depend a priori on $z$ and $\l$, are determined by the boundary conditions (\ref{eq:condlim}).
We thus obtain
\beq\label{eq:F3zlx}
\w F^{\3}(z,\l,x)=z\,\frac{\cos z(\l-x)-\sin zx}{1-\sin z\l}.
\eeq
Hence
\beqa\label{eq:Q3z}
\w Q^{\3}(z)&=&\sum_{n\ge2}z^n Q^{\3}_n
=z\int_0^1 \dd\l\int_0^\l{\rm d}x\,\w F^{\3}(z,\l,x)
\nonumber\\
&=&-z-\ln(1-\sin z),
\eeqa
and
\beqa\label{eq:I3z}
\w I^{\3}(z)&=&\sum_{n\ge1}z^n I^{\3}_n
=-1+\frac{1}{z}\int_0^1 \dd\l\,\w F^{\3}(z,\l,0)
\nonumber\\
&=&-1-\frac{1}{z}\,\ln(1-\sin z)
=\frac{\w Q^{\3}(z)}{z}.
\eeqa

\medskip
\noindent{\it Relationship with Euler numbers.}
Consider $n$ positive numbers $x_1,\dots,x_{n}$ such that
$x_i+x_{i+1}\le1$ for $1\le i\le n-1$.
These conditions define a volume $V_n$ for every integer~$n$.
The generating function of these numbers reads \cite{stanley}
\bea
\w V(z)&=&\sum_{n\ge0}z^nV_n=\frac{1}{\cos z}+\tan z
\\
&=&1+z+\frac{z^2}{2}+\frac{z^3}{3}+\frac{5}{24}z^4+\cdots
\eea
We have
\be
V_n=\frac{E_n}{n!},
\ee
where $E_n=1,1,1,2,5,16,61,\dots$ are the Euler numbers,
listed as sequence number A000111
in the On-Line Encyclopedia of Integer Sequences \cite{OEIS}.
The volumes $V_n$ are also simply related to the $Q^{\3}(n)$, as we now show.
Let us note (see~(\ref{eq:Q3z})) that
\be
\frac{\dd\w Q^{\3}(z)}{{\rm d}z}=\w V(z)-1,
\ee
hence, for $n\ge2$,
\be
Q^{\3}_n=\frac{V_{n-1}}{n}=\frac{E_{n-1}}{n!}.
\ee
The relationship between the two sequences $Q^{\3}_n$ and $V_n$ comes from the fact that
\bea
\int_0^{\l}{\rm d}x' F_{n-1}^{\3}(\l,x')&=&\prob(L_{n-1}<\l)
\\
&=&\prob(Y_1<\l,\dots,Y_{n-1}<\l)
=V_{n-1}\,\l^{n-1},
\eea
hence, integrating over $\l$,
\be
Q^{\3}_n=\int_0^1\dd\l\, V_{n-1}\,\l^{n-1}=\frac{V_{n-1}}{n}.
\ee
Let us remark that
\be
\w F^{\3}(z,\l,x)=z\left(\w V(z\l)\cos zx-\sin zx\right).
\ee
Finally, the generating function $\w V(z)$ has a pole at $z=\pi/2$, with residue 2, and therefore
\be
V_n\approx2\left(\frac{2}{\pi}\right)^{n+1},\quad
Q_n^\3\approx\frac{2}{n}\left(\frac{2}{\pi}\right)^n,\quad
I_n^\3\approx\frac{4}{\pi n}\left(\frac{2}{\pi}\right)^n.
\ee

\subsection{Analysis of sector~\2}

The generating function
\be
\w F^{\2}(z,\l,x)=\sum_{n\ge1}z^nF^{\2}_n(\l,x),
\ee
satisfies
\be
\w F'^{\2}(z,\l,x)=-z\w F^{\2}(z,\l,\l-x),
\ee
because of~(\ref{eq:derivUnif1}),
\be
\w F''^{\2}(z,\l,x)=-z^2\w F^{\2}(z,\l,x),
\ee
because of~(\ref{eq:derivUnif2}), and
\be
\w F^{\2}(z,\l,1)=z\left(1+(\l-1)\w F^{\2}(z,\l,\l-1)\right),
\ee
as a consequence of~(\ref{eq:unifF123}),
using the fact that $F_n^\1(\l,x)$ is independent of $x$.

We thus obtain, in analogy with~(\ref{eq:F3zlx})
\beq\label{eq:F2zlx}
\w F^{\2}(z,\l,x)
=z\,\frac{\cos z(\l-x)-\sin zx}{\Delta(z,\l)},
\eeq
with
\be
\Delta(z,\l)=z(\l-1)(\sin z(\l-1)-\cos z)+\cos z(\l-1)-\sin z.
\ee
We have therefore
\beqa\label{eq:Q2z}
\w Q^{\2}(z)&=&\sum_{n\ge2}z^n Q^{\2}_n
=z\int_1^2\dd\l\int_{\l-1}^1{\rm d}x\,\w F^{\2}(z,\l,x)
\nonumber\\
&=&z\int_1^2\dd\l
\,\frac{\cos z+\sin z-\cos z(\l-1)-\sin z(\l-1)}{\Delta(z,\l)},
\eeqa
and
\beqa\label{eq:I2z}
\w I^{\2}(z)&=&\sum_{n\ge1}z^n I^{\2}_n
=-1+\frac{1}{z}\int_1^2\dd\l\,\w F^{\2}(z,\l,\l-1)
\nonumber\\
&=&-1+\frac{1}{z}\int_1^2\dd\l\,
\frac{\cos z-\sin z(\l-1)}{\Delta(z,\l)}.
\eeqa
At variance with~(\ref{eq:Q3z}) and~(\ref{eq:I3z}),
the integrals over $\l$ in (\ref{eq:Q2z}) and (\ref{eq:I2z}) cannot be carried out in closed form.

\subsection{Results}

By expanding the integrands of~(\ref{eq:Q2z}) and~(\ref{eq:I2z})
as power series in $z$, integrating over~$\l$ term by term,
and adding up the contributions of~(\ref{eq:Q3z}) and~(\ref{eq:I3z}),
we obtain exact rational expressions for the probability of record breaking $Q_n$,
the mean number of records $\mean{M_n}$ and the mean value of the maximum $\mean{L_n}$.
These outcomes are given in table~\ref{tab:uni} up to $n=8$.

\begin{table}[!ht]
\begin{center}
\begin{tabular}{|c|c|c|c|c|c|c|c|}
\hline
$n$ & 2 & 3 & 4 & 5 & 6 & 7 & 8\\
\hline
$Q_n$ & 1 & $\frad{1^{\vphantom{M}}}{2_{\vphantom{M}}}$ &
$\frad{3}{8}$ & $\frad{17}{60}$ & $\frad{11}{48}$ & $\frad{481}{2520}$ & $\frad{439}{2688}$\\
\hline
$\mean{M_n}$ & 1 & $\frad{3^{\vphantom{M}}}{2_{\vphantom{M}}}$ &
$\frad{15}{8}$ & $\frad{259}{120}$ & $\frad{191}{180}$ & $\frad{2599}{1008}$ & $\frad{22109}{8064}$\\
\hline
$\mean{L_n}$ & 1 & $\frad{7^{\vphantom{M}}}{6_{\vphantom{M}}}$ &
$\frad{77}{60}$ & $\frad{49}{36}$ & $\frad{511}{360}$ & $\frad{3691}{2520}$ & $\frad{272369}{181440}$\\
\hline
\end{tabular}
\caption
{Exact values of the probability of record breaking $Q_n$,
mean number of records $\mean{M_n}$ and mean maximum $\mean{L_n}$
up to $n=8$, for a uniform distribution.}
\label{tab:uni}
\end{center}
\end{table}

The asymptotic behavior at large $n$ of the various quantities of interest
can be derived as follows.
First of all, the contribution of sector~\3 is exponentially small,
and therefore entirely negligible.
Setting again $z=\e^{-\eps}$, the integrals
entering~(\ref{eq:Q2z}) and~(\ref{eq:I2z}) are dominated
by a range of values of the difference $2-\l$ that shrinks proportionally to $\sqrt{\eps}$ as $\eps\to0$.
Changing integration variable from $\l$ to $t$ such that $\l=2-t\sqrt{\eps}$,
and keeping only terms which are singular in $\eps$, we obtain
\bea
\w Q^\2(z)&=&
\ln\frac{1}{\eps}\left(1-\frac{\eps}{3}+\cdots\right)
-\pi\sqrt\frac{\eps}{2}\left(1-\frac{5\eps}{9}+\cdots\right),
\\
\w I^\2(z)&=&
\frac{2}{3}\ln\frac{1}{\eps}\left(1+\frac{38\eps}{45}+\cdots\right)
+\frac{\pi}{\sqrt{2\eps}}\left(1+\frac{5\eps}{6}+\cdots\right),
\eea
and so
\beqa\label{eq:uniasy}
nQ_n&=&1+\delta_n
\nonumber\\
&=&1+\frac{1}{3n}+\cdots
+\sqrt\frac{\pi}{8n}\left(1+\frac{5}{6n}+\cdots\right),
\eeqa
and
\bea
\mean{L_n}&=&
2\left(1-\frac{1}{3n}+\cdots\right)
-\sqrt\frac{\pi}{2n}\left(1-\frac{5}{12n}+\cdots\right).
\eea
Finally, omitting details,
we obtain a similar asymptotic expansion for the mean number of records, i.e.,
\be
\mean{M_n}=\ln n+K+\frac{1}{6n}+\cdots
-\sqrt\frac{\pi}{2n}\left(1+\frac{1}{36n}+\cdots\right),
\ee
where the finite part reads
\bea
K&=&
\gamma-1-\ln(2(1-\sin 1))+2\int_1^2\dd\l\left(\!\frac{1}{(2-\l)\cot\frac{2-\l}{2}-\l}-\frac{1}{2-\l}\!\right)
\nonumber\\
&=&1.092998\dots
\eea

\section{Power-law distribution with index $\th=1$}
\label{sec:th1}

The case where the random variables $X_i$ have a power-law distribution with index $\th=1$,
with common density $f(x)=1/x^2$ and distribution function $F(x)=1-1/x$ for $x>1$ (see table~\ref{tab:distribs}),
is our last example giving rise to an exact solution of the problem,
although end results are somewhat less explicit than in the two previous cases.
The distribution under consideration is marginal,
in the sense that $\mean{X}$ is logarithmically divergent.

\subsection{Differential equations}

In the present case, the key equation~(\ref{eq:start}) reads
\be
F'_n(\l,x)=-\frac{2}{x}F_n(\l,x)-\frac{1}{x^2}F_{n-1}(\l,\l-x)
\ee
for $n\ge2$, $\l>2$, and $1<x<\l-1$.
Setting
\beq\label{FtoH}
F_n(\l,x)=\frac{H_n(\l,x)}{x^2(\l-x)},
\eeq
the new functions $H_n(\l,x)$ obey the differential equation
\beq\label{H1dif}
x(\l-x)H'_n(\l,x)+xH_n(\l,x)=-H_{n-1}(\l,\l-x),
\eeq
with boundary condition $H_n(\l,\l-1)=0$,
as well as
\beq\label{H2dif}
x^2(\l-x)^2H''_n(\l,x)=-H_{n-2}(\l,x).
\eeq
We thus obtain
\bea
H_1(\l,x)&=&\l-x,
\\
H_2(\l,x)&=&\l-x-1,
\\
H_3(\l,x)&=&\frac{(\l-1)(\l-1-x)}{\l}+\frac{\l-x}{\l^2}\ln\frac{x}{(\l-1)(\l-x)}.
\eea

\subsection{Generating function}

In order to solve the recursive differential equations~(\ref{H1dif}),~(\ref{H2dif}),
we introduce the generating function
\beq\label{Hgene}
\w H(z,\l,x)=\sum_{n\ge1}z^nH_{n}(\l,x),
\eeq
which obeys
\beq\label{Hdif1}
x(\l-x)\w H'(\l,x)+x\w H(\l,x)=-z\w H(\l,\l-x),
\eeq
with boundary condition
\beq\label{Hbc}
\w H(\l,\l-1)=z,
\eeq
as well as
\beq\label{Hdif2}
x^2(\l-x)^2\w H''(\l,x)=-z^2\w H(\l,x).
\eeq
The general solution to~(\ref{Hdif2}) reads
\be
\w H(z,\l,x)=A_+
x^{a_+}(\l-x)^{a_-}
+A_-
x^{a_-}(\l-x)^{a_+},
\ee
with
\be
a_{\pm}=\frac{1\pm w}{2},\quad w=\sqrt{1-\frac{4z^2}{\l^2}}.
\ee
Notice the similarity with~(\ref{fgeneres}).
The amplitudes $A_{\pm}$ are determined by~(\ref{Hdif1}) and~(\ref{Hbc}), yielding
\beq\label{Hfull}
\w H(z,\l,x)=z\,\frac
{\sqrt{a_-}\,x^{a_+}(\l-x)^{a_-}-\sqrt{a_+}\,x^{a_-}(\l-x)^{a_+}}
{\sqrt{a_-}\,(\l-1)^{a_+}-\sqrt{a_+}\,(\l-1)^{a_-}}.
\eeq
This result demonstrates that the functions $H_n(\l,x)$
only involve integer powers of $\ln x$ and $\ln(\l-x)$, besides rational functions.

\subsection{Probability of record breaking}

The generating function of the $Q_n$ reads
\be
\w Q(z)=\sum_{n\ge2}z^{n}Q_n=z\int_2^\infty\dd\l
\int_1^{\l-1}\frac{{\rm d}x}{x^2(\l-x)^3}\,\w H(z,\l,x),
\ee
by virtue of~(\ref{eq:Qn}),~(\ref{FtoH}) and~(\ref{Hgene}),
where $\w H(z,\l,x)$ is given by~(\ref{Hfull}).
The integral over $x$ can be carried out in closed form.
We thus obtain
\beq\label{Qzinv}
\w Q(z)=\int_2^\infty\dd\l\,I(z,\l),
\eeq
with
\be
I(z,\l)=z^2\,\frac
{\sqrt{a_-}\,I_+(z,\l)-\sqrt{a_+}\,I_-(z,\l)}
{\sqrt{a_-}\,(\l-1)^{a_+}-\sqrt{a_+}\,(\l-1)^{a_-}}
\ee
and
\bea
I_\pm(z,\l)
&=&\int_1^{\l-1}{\rm d}x\,\frac{x^{a_\pm}(\l-x)^{a_\mp}}{x^2(\l-x)^3}
\\
&=&\int_1^{\l-1}{\rm d}x\,x^{-2+a_\pm}(\l-x)^{-2-a_\pm}
\\
&=&\frac{1}{(1-a_\pm)\l^3}\left((\l-1)^{1-a_\pm}-(\l-1)^{-(1-a_\pm)}\right)
\\
&+&\frac{1}{(1+a_\pm)\l^3}\left((\l-1)^{1+a_\pm}-(\l-1)^{-(1+a_\pm)}\right)
\\
&+&\frac{2}{a_\pm\l^3}\left((\l-1)^{a_\pm}-(\l-1)^{-a_\pm}\right).
\eea

As was the case for~(\ref{Qzexp}) and~(\ref{eq:Q2z}),
the integrals over $\l$ in~(\ref{Qzinv}) cannot be carried out analytically in closed form.
By expanding the integrand in~(\ref{Qzinv}) as a power series in $z$
and integrating term by term with respect to~$\l$,
we obtain the following values for the probability of record breaking $Q_n$,
besides the universal ones derived in section~\ref{sec:firstn}:
\bea
Q_5&=&\frac{5}{8}-\frac{\pi^2}{30}=0.296013\dots,
\nonumber\\
Q_6&=&\frac{61}{144}+\frac{14\pi^2}{135}-\zeta(3)=0.245068\dots,
\nonumber\\
Q_7&=&\frac{475}{252}-\frac{292\pi^2}{945}+\frac{8\zeta(3)}{7}=0.209044\dots,
\eea
and so on.
In contrast with the two previous exactly solvable cases
(see tables~\ref{tab:exp} and~\ref{tab:uni}),
here the non-universal $Q_n$ are not rational,
and they involve the values of Riemann zeta function at larger and larger positive integers.

The asymptotic behavior of $Q_n$ at large $n$ can be derived from~(\ref{Qzinv})
by setting again $z=\e^{-\eps}$,
and considering the regime where $\eps$ and $1/\l$ are simultaneously small.
To leading order,~(\ref{Qzinv}) reduces to
\be
\w Q(z)\approx\int_2^\infty\dd\l\,\frac{3}{2\l(1+\eps \l)},
\ee
i.e., performing the inverse Laplace transform,
\be
Q_n\approx\int_2^\infty\dd\l\,\frac{3}{2\l^2}\,\e^{-n/\l}\approx\frac{3}{2n},
\ee
up to negligible boundary terms.
The above result is an explicit instance where (\ref{eq:lim32}) holds.
A full asymptotic expansion of $Q_n$ can be derived by keeping track of higher orders, yielding
\beq\label{eq:Qinvasy}
nQ_n=1+\delta_n=\frac{3}{2}-\frac{2(\ln n+\gamma-3)}{n}+\cdots
\eeq

Finally, omitting details,
we obtain a similar asymptotic expansion for the mean number of records, i.e.,
\be
\mean{M_n}=\frac{3}{2}\,(\ln n+K)+\frac{2(\ln n+\gamma-2)}{n}+\cdots,
\ee
where the finite part reads
\be
K=\gamma-\ln 2+\int_2^\infty\dd\l\left(\frac{2I(1,\l)}{3}-\frac{1}{\ell}\right)
=-0.387293\dots
\ee

\subsection{Distribution of the maximum}

Here $\mean{X}$ is divergent,
so that it makes no sense to evaluate $\mean{L_n}$.
The full distribution of $L_n$ should be considered instead.
We have
\be
\F_n(\l)=\int_1^{\l-1}{\rm d}x\,F_n(\l,x)
=F_{n+1}(\l,1)=\frac{H_{n+1}(\l,1)}{\l-1},
\ee
as a consequence of~(\ref{FLint}),~(\ref{eq:recurF}) and~(\ref{FtoH}).
We thus obtain
\bea
\F_1(\l)&=&1-\frac{1}{\l-1},
\\
\F_2(\l)&=&1-\frac{2}{\l}-\frac{2}{\l^2}\,\ln(\l-1),
\\
\F_3(\l)&=&1-\frac{3(\l-1)}{\l^2}-\frac{1}{\l^2(\l-1)}-\frac{4(\l-1)}{\l^3}\,\ln(\l-1).
\eea

The corresponding generating function reads
\bea
\w {\cal F }(z,\l)&=&\sum_{n\ge0}z^n\F_n(\l)=\frac{\w H(z,\l,1)}{z(\l-1)}
\nonumber\\
&=&\frac
{\sqrt{a_-}\,(\l-1)^{-a_+}-\sqrt{a_+}\,(\l-1)^{-a_-}}
{\sqrt{a_-}\,(\l-1)^{a_+}-\sqrt{a_+}\,(\l-1)^{a_-}},
\eea
where the last expression is a consequence of~(\ref{Hfull}).

The scaling behavior of the distribution of $L_n$ at large $n$
can be derived along the lines of the previous section.
To leading order, we find the simple result
\beq\label{eq:frechet}
\F_n(\l)\approx\e^{-n/\l}.
\eeq
In particular, the median value $L_n^\star$, such that $\F_n(L_n^\star)=\frac12$,
reads
\be
L_n^\star\approx\frac{n}{\ln 2}.
\ee
The full asymptotic expansion of $\F_n(\l)$ in the regime where $n$ and $\l$ are comparable reads
\beq\label{Finvasy}
\F_n(\l)=\e^{-n/\l}\left(1-\frac{n}{2\l^2}(4\ln\l-1)+\cdots\right),
\eeq
and so
\be
L_n^\star=\frac{n}{\ln 2}+2\ln\frac{n}{\ln 2}-\frac{1}{2}+\cdots
\ee

\section{Asymptotic analysis of the general case}
\label{sec:gal}

The probability of record breaking $Q_n$
exhibits a great variety of asymptotic behaviors,
depending on the underlying distribution $f(x)$.
This is exemplified by the three exactly solvable cases
studied in sections~\ref{sec:exp} to~\ref{sec:th1}.
In terms of the correction $\delta_n$ such that $nQ_n=1+\delta_n$
(see~(\ref{eq:deldef})),
we have seen that
\beq\label{delmargin}
\delta_n\approx\frac{1}{\ln n}
\eeq
for the exponential distribution (see~(\ref{eq:lexpasy})),
\be
\delta_n\approx\sqrt\frac{\pi}{8n}
\ee
for the uniform distribution (see~(\ref{eq:uniasy})),
and
\beq\label{eq:deluni}
\delta_n\to\frac{1}{2},
\eeq
which is equivalent to~(\ref{eq:lim32}),
for the power-law distribution with $\th=1$ (see~(\ref{eq:Qinvasy})).

This section is devoted to a heuristic but systematic analysis
of the dependence of the asymptotic behavior of $\delta_n$
on the underlying distribution $f(x)$.
It will turn out that the exponential distribution,
where $\delta_n$ falls off logarithmically (see~(\ref{delmargin})),
is a marginal case.
For superexponential distributions,
the analysis of sections~\ref{sec:predict} and \ref{sec:es}
demonstrates that $\delta_n$ falls off to zero
and yields a general asymptotic formula for $\delta_n$ (see~(\ref{eq:delheu})).
For subexponential distributions,
it will be shown in section~\ref{sec:se}
that $nQ_n$ and~$\delta_n$ go to the universal limits~(\ref{eq:lim32}) and~(\ref{eq:deluni}).
This dichotomy will be extended to higher values of the window width $p$ in section~\ref{sec:p}.

\subsection{Cyclization of the sequence}

The first step of the analysis consists in comparing the problem at hand
with a cyclic variant of it.
For the former, we have
\be
Q_n=\prob(Y_n>L_{n-1}),
\ee
with
\be
L_n=\max(Y_2,\dots,Y_n)
\ee
(see~(\ref{eq:Qndef}) and~(\ref{eq:Ln})).
The cyclic variant of the problem is defined by introducing
\be
Y_1^\cy=X_n+X_1.
\ee
The sequence $Y_1^\cy,Y_2,\dots,Y_n$ thus obtained
involves the basic variables $X_1,\dots,X_n$
in a cyclically invariant fashion.
It has therefore exchangeable entries, and~so
\be
Q_n^\cy=\prob\big(Y_n>\max(Y_1^\cy,Y_2,\dots,Y_{n-1})\big)=\frac{1}{n}.
\ee
Introducing the events
\bea
E=\{Y_n>Y_1^\cy\}=\{X_{n-1}>X_1\},
\nonumber\\
F=\{Y_n>L_{n-1}\}=\{L_n=Y_n\},
\eea
we have
\bea
\prob(E\cap F)=Q_n^\cy=\frac{1}{n},
\nonumber\\
\prob(F)=Q_n=\frac{1+\delta_n}{n},
\eea
and so
\beqa\label{eq:deltan}
\Delta_n&=&\prob(\bar E|F)=\prob(X_1>X_{n-1}|L_n=Y_n)
\nonumber\\
&=&\frac{Q_n-Q_n^\cy}{Q_n}=\frac{\delta_n}{1+\delta_n}.
\eeqa
This equation gives a description of the difference $Q_n-Q_n^\cy$
in terms of a conditional probability,
which will prove useful in the following.

\subsection{Decoupled model}

We now consider a decoupled variant of the original problem,
whose main advantage is that the expression (\ref{eq:deltan})
can be given the explicit form~(\ref{eq:Deltaheu}),
which will in turn yield the estimate~(\ref{eq:delheu}) for the correction $\delta_n$
in appropriate situations.

The decoupled model is defined as follows.
The random variables $Y_i$ of the original problem are replaced by
a sequence of iid random variables
\beq\label{eq:Ycal}
\Y_i=X_i+X'_i,
\eeq
where $X_i$ and $X'_i$ are two independent replicas of the original random variables $X_i$
with common density $f(x)$ and distribution function $F(x)$.
The number of variables~$X$ is therefore doubled with respect
to the original problem.
The distribution function~$F_2(y)$ and the density $f_2(y)$
of the variables $\Y_i$ thus read
\beqa\label{eq:convol2}
F_2(y)&=&\prob(\Y<y)=\prob(X+X'<y)=\int_0^y{\rm d}y' f_2(y'),
\nonumber\\
f_2(y)&=&\int_0^y{\rm d}x\,f(x) f(y-x).
\eeqa
In terms of the Laplace transform
\be
\hat f(s)=\int_0^\infty {\rm d}x\,\e^{-sx}\,f(x),
\ee
this reads
\beq\label{eq:Lapconvol}
\hat f_2(s)=\hat f(s)^2.
\eeq
The conditional density of $X$ given $X+X'=y$, denoted by $f(x|y)$, is equal to
\beq\label{eq:fxycond}
f(x|y)=\frac{f(x)f(y-x)}{f_2(y)}.
\eeq
The largest among the first $n$ variables $\Y_i$, denoted by
\be
\Y^*=X^*+X'^*,
\ee
has distribution function
\be
F_{\Y^*}(y)=\prob(\Y^*<y)=F_2(y)^n,
\ee
and density
\beq\label{eq:fystar}
f_{\Y^*}(y)=nF_2(y)^{n-1}f_2(y).
\eeq
Using~(\ref{eq:fxycond}) and~(\ref{eq:fystar}), the density of $X^*$ is
\be
f_{X^*}(x)
=\int_0^\infty{\rm d}y\,f(x|y) f_{\Y^*}(y)
=n f(x)\int_x^\infty{\rm d}y\,f(y-x) F_2(y)^{n-1}.
\ee

Within the setting of the decoupled model,
the conditional probability $\Delta_n$ introduced in~(\ref{eq:deltan}) therefore reads
\beqa\label{eq:Deltaheu}
\Delta_n
&=&\prob(X>X^*)=\int_0^\infty{\rm d}x\,f_{X^*}(x)\bar F(x)
\nonumber\\
&=&n\int_0^\infty{\rm d}y\,F_2(y)^{n-1}
\int_0^y{\rm d}x\underbrace{f(x)f(y-x)}\bar F(x),
\eeqa
with
\be
\bar F(x)=\prob(X>x)=1-F(x).
\ee
When~$n$ is large,
the factor $F_2(y)^{n-1}$ in (\ref{eq:Deltaheu}) selects large values of $y$,
such that $\bar F_2(y)$ scales as $1/n$.
These are the typical values of $\Y^*$.
The product underlined by the brace,
which already entered~(\ref{eq:convol2}) and~(\ref{eq:fxycond}),
describes to what extent the distribution of~$X$ is affected by the conditioning
by such a large value $y$ of the sum $\Y=X+X'$.

\subsection{The key dichotomy}
\label{sec:predict}

The dichotomy between the two limits~(\ref{eq:lim1}) and~(\ref{eq:lim32})
is now shown in gene\-ral albeit non-rigorous terms
to be dictated by the form of the tail of the under\-lying parent distribution~$f(x)$
or, equivalently, by the analytic structure of its Laplace transform~$\hat f(s)$.

\begin{enumerate}
\item[$\circ$]
For {\it superexponential distributions},
i.e., distributions $f(x)$ either having a bounded support
or falling off faster than any exponential,
such as e.g.~a half-Gaussian or any other compressed exponential,
$\hat f(s)$ is an entire function, i.e., it is analytic in the whole complex $s$-plane.
Then, as a general rule, $f_2(y)$ (see (\ref{eq:convol2})) has a slower decay than~$f(x)$.
Furthermore, if the sum $\Y=X+X'$ is atypically large,
then both $X$ and $X'$ are atypically large as well, with very high probability.
As a consequence, the conditional probability~$\Delta_n$,
as given by~(\ref{eq:Deltaheu}), falls off to zero for large~$n$.
Simplifying the latter expression,
we thus obtain the following asymptotic estimate for~$\delta_n$:
\beq\label{eq:delheu}
\delta_n\approx
n\int_0^\infty{\rm d}y\,\e^{-n\bar F_2(y)}\int_0^y{\rm d}x\,f(x)f(y-x)\bar F(x).
\eeq
We claim that this prediction becomes asymptotically exact for all superexponential distributions,
in the sense that it correctly describes the decay of $\delta_n$, to leading order for large $n$,
in spite of its heuristic derivation using the decoupled model.
The rationale behind this claim
is that the difference between the original and the decoupled models,
measured by the relative difference between $Q_n$ and $Q_n^\cy$,
is consistently found to decay to zero, proportionally to the estimate~(\ref{eq:delheu}) for~$\delta_n$.

\item[$\circ$]
For {\it subexponential distributions},
i.e., distributions $f(x)$ which fall off smoothly enough and less rapidly than any exponential,
such as e.g.~a power law or a stretched exponential,
$\hat f(s)$ has an isolated branch-point singularity at $s=0$.
The asymptotic equivalence of the tails,
\beq\label{eq:subexp}
\bar F_2(y)\approx 2\bar F(y)\quad(y\gg1),
\eeq
can be derived by an inverse Laplace transform of (\ref{eq:Lapconvol}),
where the contour integral is dominated by the singularity of $\hat f(s)$ at $s=0$.
Equation~(\ref{eq:subexp}) may be used as a mathematically rigorous definition
of the class of subexponential distributions,
following Chistyakov \cite{chistyakov}.
Its intuitive meaning is the following:
if the sum $\Y=X+X'$ is very large,
then one of the terms, either $X$ or~$X'$---hence the factor 2---is typical,
i.e., distributed according to $f(x)$,
while the other one is essentially equal to $\Y$.
This behavior underlies the phenomenon of condensation for subexponential random variables
conditioned by an atypical value of their sum (see \cite{cg} for a recent review and the references therein).
As a consequence of (\ref{eq:subexp}), for subexponential distributions~$f(x)$,
the estimate~(\ref{eq:Deltaheu}) remains of order unity for large $n$.
The decoupled model is therefore of little use to understand the original one.
This situation will be investigated in section~\ref{sec:se},
where $nQ_n$ and~$\delta_n$ will be shown to admit the universal limits~(\ref{eq:lim32}) and~(\ref{eq:deluni}).

\end{enumerate}

For {\it exponential distributions},
i.e., distributions $f(x)$ falling off either as a pure exponential $\e^{-\beta x}$, with $\beta>0$,
or as the product of such an exponential by a more slowly varying prefactor,
such as e.g.~a power of $x$,
the leading (i.e., rightmost) singularity of $\hat f(s)$ is located on the negative real axis at $s=-\beta$.
For our purpose, these distributions are marginal
since they can lie on either sides of the dichotomy between~(\ref{eq:lim1}) and~(\ref{eq:lim32})
(see section~\ref{sec:expui}).

\subsection{Superexponential and (some) exponential distributions}
\label{sec:es}

The prediction~(\ref{eq:delheu})
is now made explicit for a variety of superexponential and exponential distributions~$f(x)$.

\subsubsection{Pure exponential distribution.}
This is the distribution for which an exact solution has been presented in section~\ref{sec:exp}.
We have
\be
f(x)=\bar F(x)=\e^{-x},\quad
f_2(y)=y\,\e^{-y},\quad\bar F_2(y)=(y+1)\e^{-y}.
\ee
The estimate~(\ref{eq:delheu}) therefore reads
\beq\label{eq:delnint}
\delta_n\approx n\int_0^\infty{\rm d}y\,\exp\left(-y-n(y+1)\e^{-y}\right).
\eeq
This integral can be evaluated in analogy with~(\ref{Qnint}).
Setting $\lambda=\ln n$ (see~(\ref{lamdef})) and $(y+1)\e^{-y}=\e^{-\mu}$, we obtain
\be
\delta_n\approx\int_{-\infty}^\infty
\dd\mu\,\exp\left(\lambda-\mu-\e^{\lambda-\mu}\right)\frac{1}{y(\mu)},
\ee
hence
\beq\label{eq:dexpheu}
\delta_n\approx\frac{1}{\lambda}-\frac{\ln\lambda+\gamma}{\lambda^2}+\cdots
\eeq
A comparison with the exact expansion~(\ref{eq:lexpasy})
shows that the estimate~(\ref{eq:delheu}) is correct to leading order in this marginal case.
The difference between the estimate (\ref{eq:dexpheu}) and the exact result
is indeed subleading, since it scales as $2/\lambda^2$.

\subsubsection{Exponential distribution modulated by a power law.}
\label{sec:expui}

We now consider distri\-butions falling off as an exponential modulated by a power law, i.e.,
\beq
f(x)\approx\bar F(x)\approx A\,x^{a-1}\,\e^{-x}\quad(x\to\infty),
\label{expui}
\eeq
where $a$ is arbitrary (positive or negative).

Let us consider first the case where $a>0$.
We have then
\be
\hat f(s)\approx\frac{A\Gamma(a)}{(s+1)^a}\quad(s\to-1)
\ee
and
\beq
f_2(y)\approx\bar F_2(y)\approx B\,y^{2a-1}\,\e^{-y}\quad(y\to\infty),
\label{expf2}
\eeq
with $B=(A\Gamma(a))^2/\Gamma(2a)$.
Performing the integrals entering~(\ref{eq:delheu}), we obtain
\be
\delta_n\approx nA^3\Gamma(a)\int_0^\infty{\rm d}y\,\exp\left(-nBy^{2a-1}\e^{-y}\right)y^{a-1}\e^{-y}.
\ee
This integral can be evaluated in analogy with~(\ref{Qnint}).
Setting $\lambda=\ln (nB)$ and $y^{2a-1}\e^{-y}=\e^{-\mu}$, we obtain formally
\be
\delta_n\approx\frac{A^3\Gamma(a)}{B}
\int_{-\infty}^\infty\dd\mu\,\exp\left(\lambda-\mu-\e^{\lambda-\mu}\right)\frac{1}{y(\mu)^a}.
\ee
To leading order, the identification $y(\mu)\approx\mu\approx\lambda$ yields the estimate
\beq
\delta_n\approx\frac{A\Gamma(2a)}{\Gamma(a)}\,\frac{1}{(\ln n)^a}.
\label{delexpui}
\eeq

We are thus led to claim that exponential distributions of the form (\ref{expui}) with $a>0$,
and presumably all exponential distributions such that $\hat f(s)\to+\infty$
as the leading singularity is approached from the right ($s\to-\beta^+$),
belong to the superexponential side of the dichotomy, in the sense that (\ref{eq:lim1}) holds,
and that~(\ref{delexpui}) correctly predicts the decay of the correction~$\delta_n$.
The logarithmically slow fall off of the latter expression
confirms the marginal character of this class of exponential distributions.

On the contrary, if the exponent $a$ entering (\ref{expui}) is negative,
the above derivation already breaks down at the level of~(\ref{expf2}).
Exponential distributions of the form (\ref{expui}) with $a<0$,
and presumably all exponential distributions
such that~$\hat f(s)$ remains bounded as $s\to-\beta^+$,
therefore share with subexponential distributions the property that the estimate $\delta_n$ does not decay to zero,
with the expected consequence that (\ref{eq:lim32}) should hold.

\subsubsection{Distributions with bounded support and power-law singularity.}

We now consider the case where $f(x)$ is supported by the interval [0, 1]
and has a power-law singularity at its upper edge,~i.e.,
\beqa\label{eq:power}
\bar F(x)\approx A\eps^a,\quad f(x)\approx aA\eps^{a-1},
\nonumber\\
\bar F_2(y)\approx B\eta^{2a},\quad f_2(y)\approx 2aB\eta^{2a-1},
\eeqa
with the notations $\eps=1-x$, $\eta=2-y$.
The exponent $a>0$ and the ampli\-tude $A>0$ are arbitrary.
We have $B=a(A\Gamma(a))^2/(2\Gamma(2a))$.
Performing the integrals entering~(\ref{eq:delheu}),
we obtain a universal $1/\sqrt{n}$ decay for $\delta_n$,
irrespective of the exponent~$a$,~i.e.,
\beq\label{eq:racinv}
\delta_n\approx\frac{K(a)}{\sqrt{n}},
\eeq
where the amplitude $K(a)$ reads
\beq\label{eq:kofa}
K(a)=\frac{1}{\Gamma(a)^2\,\Gamma(3a)}\sqrt{\frac{\pi\Gamma(2a)^5}{2a}}.
\eeq

\begin{figure}[!ht]
\begin{center}
\includegraphics[angle=0,width=.7\linewidth]{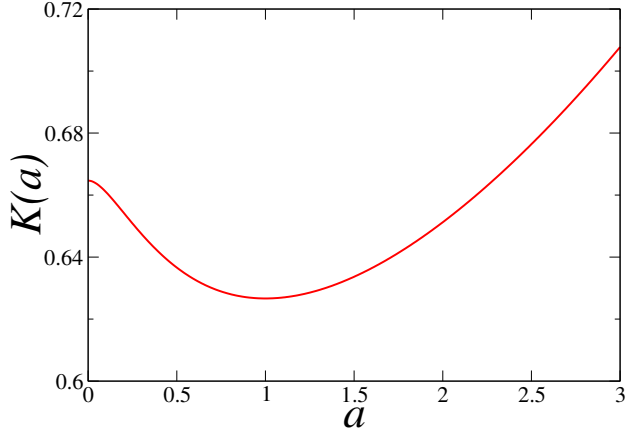}
\caption
{Amplitude $K(a)$ of the universal $1/\sqrt{n}$ decay~(\ref{eq:racinv})
of the correction term $\delta_n$, against the exponent $a$.}
\label{fig:kofa}
\end{center}
\end{figure}

The amplitude $K(a)$ is shown in figure~\ref{fig:kofa}.
It has a local maximum at $K(0)=3\sqrt{\pi}/8=0.664670\dots$
and a local minimum at $K(1)=\sqrt{\pi/8}=0.626657\dots$
The latter value agrees with the exact result~(\ref{eq:uniasy}) for the uniform distribution.
This provides another corroboration
of our claim that the estimate~(\ref{eq:delheu}) is correct to leading order.
The exponential growth $K(a)\sim(32/27)^a$ of the amplitude at large $a$
suggests that the $1/\sqrt{n}$ decay
ceases to hold for distributions with an infinitely large exponent,
i.e., with an essential singularity at their upper edge.

\subsubsection{Distributions with bounded support and exponential singularity.}

We now consider the case where $f(x)$ is supported by the interval [0, 1]
and has an exponentially small singularity at its upper edge, of the form
\beq\label{bse}
f(x)\sim\bar F(x)\sim\e^{-C/\eps^b},
\eeq
with $b>0$.
Using the same notations as above, and working within exponential accuracy,
we have
\be
f_2\sim\int_0^\eta\dd\eps\,\e^{-C(1/\eps^b+1/(\eta-\eps)^b)},
\ee
for small $\eta$,
where the integral is dominated by a saddle point at $\eps=\eta/2$, so that
\be
f_2\sim\bar F_2\sim\e^{-2^{1+b}C/\eta^b}.
\ee
Similarly, the $x$-integral entering~(\ref{eq:delheu}) is dominated
by a saddle point at $\eps=\tau\eta$, with $\tau=1/(1+2^{-1/(b+1)})$, and so
\be
\delta_n\sim\int_0^\infty\dd\eta\,
\exp\!\left(-\frac{2C}{\tau^{b+1}\eta^b}-n\,\e^{-2^{1+b}C/\eta^b}\right).
\ee
Using once more the saddle-point method,
we obtain a power-law decay of the form
\be
\delta_n\sim n^{-\omega_1(b)},
\ee
where the exponent
\beq\label{omegab}
\omega_1(b)=\frac{(1+2^{-1/(b+1)})^{b+1}}{2^b}-1
\eeq
decreases monotonically as a function of $b$,
from $\omega_1(0)=1/2$ to $\omega_1(\infty)=\sqrt2-1$.

\subsubsection{Compressed exponential distributions.}

We now consider the case where $f(x)$ has a compressed exponential
(or superexponential) tail extending up to infinity, of the form
\beq\label{eq:compe}
f(x)\sim\bar F(x)\sim\e^{-Cx^c},
\eeq
with $c>1$.
The analysis of this case is very similar to the previous one.
We have
\be
f_2(y)\sim\int_0^y{\rm d}x\,\e^{-C(x^c+(y-x)^c)}
\ee
for large $y$,
where the integral is dominated by a saddle point at $x=y/2$, so that
\be
f_2(y)\sim\bar F_2(y)\sim\e^{-2^{1-c}Cy^c}.
\ee
Similarly, the $x$-integral entering~(\ref{eq:delheu}) is dominated
by a saddle point at $x=\tau y$, with $\tau=1/(1+2^{1/(c-1)})$, and so
\be
\delta_n\sim\int_0^\infty{\rm d}y\,\exp\!\left(-2C\tau^{c-1}y^c-n\,\e^{-2^{1-c}Cy^c}\right).
\ee
We thus obtain a power-law decay of the form
\be
\delta_n\sim n^{-\omega_2(c)},
\ee
where the exponent
\beq\label{eq:omegac}
\omega_2(c)=\frac{2^c}{(1+2^{1/(c-1)})^{c-1}}-1
\eeq
increases monotonically as a function of $c$,
from $\omega_2(c)\approx(c-1)\ln 2$ as $c\to1$ to $\omega_2(\infty)=\sqrt2-1$.
In particular, for the half-Gaussian distribution ($c=2$),
we predict the decay exponent
\beq\label{omegagau}
\omega_{\rm Gaussian}=\omega_2(2)=\frac{1}{3}.
\eeq

\begin{figure}[!ht]
\begin{center}
\includegraphics[angle=0,width=.7\linewidth]{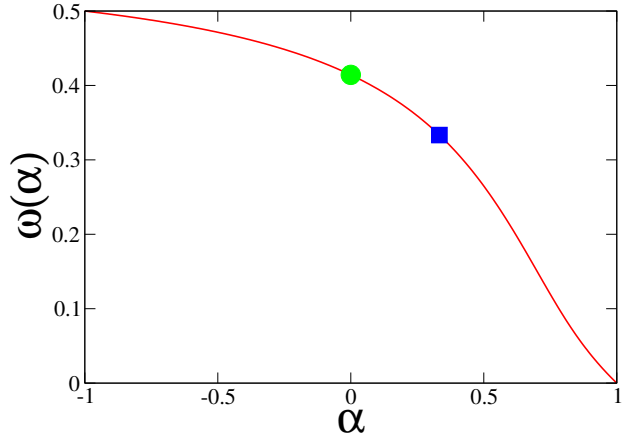}
\caption
{Exponent $\omega(\alpha)$ (see~(\ref{eq:omegatot})) unifying the decay exponents
$\omega_1(b)$ (see~(\ref{omegab})) and $\omega_2(c)$ (see~(\ref{eq:omegac}))
characteristic of distributions with exponential singularities.
Green circular symbol: limiting value $\omega(0)=\sqrt2-1$
characteristic of distributions with a double exponential fall-off.
Blue square symbol: decay exponent~(\ref{omegagau}) of the half-Gaussian distribution
($c=2$, i.e., $\alpha=1/3$).}
\label{fig:exomega}
\end{center}
\end{figure}

As it turns out,
the decay exponents $\omega_1(b)$ (see~(\ref{omegab}))
and $\omega_2(c)$ (see~(\ref{eq:omegac}))
can be unified into a single function
\beq\label{eq:omegatot}
\omega(\alpha)=2^{(\alpha+1)/(2\alpha)}
\left(1+2^{2\alpha/(1-\alpha)}\right)^{(\alpha-1)/(2\alpha)}-1
\eeq
of a parameter $\alpha$ in the range $-1<\alpha<1$,
as shown in figure~\ref{fig:exomega}.
Distributions with a bounded support and an exponential singularity with index $b$
correspond to $-1<\alpha<0$,
whereas compressed exponential distributions with index $c$
correspond to $0<\alpha<1$, with the identifications
\beq\label{bcalpha}
b=-\frac{\alpha+1}{2\alpha},\quad
c=\frac{\alpha+1}{2\alpha}.
\eeq
The exponent $\omega(\alpha)$ is a decreasing function
from $\omega(-1)=\omega_1(0)=1/2$ to $\omega(1)=\omega_2(1)=0$,
via the common limiting value $\omega(0)=\omega_1(\infty)=\omega_2(\infty)=\sqrt2-1$,
characteristic of distributions with a double exponential fall-off,
either at the upper edge of a compact support or at infinity.

\subsection{Subexponential distributions}
\label{sec:se}

We now consider subexponential distributions, whose tails decrease more slowly than any exponential.
Our goal is to show that the correction $\delta_n$
goes to the universal limit~(\ref{eq:deluni}), i.e., that $Q_n$ falls off as
\beq\label{eq:Quni}
Q_n\approx\frac{3}{2n}
\eeq
for large $n$.
This result agrees to leading order with the expansion~(\ref{eq:Qinvasy}),
ensuing from an exact solution for the power-law distribution with $\th=1$.
It also agrees with the exact expression (\ref{eq:Qeb}) of $Q_n$ for finite $n$
in the limiting situation of exponentially broad distributions.

The gist of the derivation of~(\ref{eq:Quni})
consists in looking for a solution to the integral recursion~(\ref{eq:recurF})
in an approximately factorized form, i.e.,
\beq\label{eq:Fnlxestim}
F_n(\l,x)\approx K_n\,f(x)\,(1-\eps_n(\l,x)).
\eeq
The condition~(\ref{eq:Fll}) yields
\beq\label{eq:epsbc1}
\eps_n(\l,\l)=1,
\eeq
whereas $\eps_n(\l,x)$ is assumed to be small
in the regime of interest where $n$ and $\l$ are simultaneously large,
with $x$ being kept finite.
We set
\beq\label{eq:epsbc2}
\eps_n(\l,0)=0,
\eeq
fixing thus the prefactor $K_n$ unambiguously.
To leading order, the differential equation~(\ref{eq:start}) yields
\beq\label{eq:epsdiff}
K_n\eps_n'(\l,x)\approx K_{n-1}f(\l-x).
\eeq
Equation~(\ref{eq:epsdiff}), with boundary conditions~(\ref{eq:epsbc1}) and~(\ref{eq:epsbc2}),
admits a similarity solution where $\eps(\l,x)$ and the ratio $q=K_n/K_{n-1}$
are independent of $n$, namely
\be
\eps(\l,x)=\frac{F(\l)-F(\l-x)}{F(\l)},\quad
q=F(\l).
\ee
Whenever $n$ and $\l$ are simultaneously large,
(\ref{eq:Fnlxestim}) simplifies to
\be
F_n(\l,x)\approx\e^{-n\bar F(\l)}f(x)F(\l-x),
\ee
so that~(\ref{eq:Qn}) yields the estimate
\beq\label{eq:Qndip}
Q_n\approx\int_0^\infty\dd\l\,\e^{-n\bar F(\l)}
\int_0^\l{\rm d}x\,f(x)f(\l-x)F(\l-x).
\eeq
The analysis of this expression for large $n$ is somewhat similar
to that of~(\ref{eq:Deltaheu}), performed in section~\ref{sec:es}.
The exponential factor selects large values of $\l$,
such that~$\bar F(\l)$ scales as $1/n$.
These are the typical values of $L_n$.
The subexponentiality of~$f(x)$,
in the intuitive sense explained below (\ref{eq:subexp}),
suggests that the integral over the variable~$x$ in (\ref{eq:Qndip}) is dominated by the vicinity
of its endpoints,
i.e., of the regimes where either $x$ or the difference $\l-x$ is kept finite.
Adding these two contributions yields
\be
\int_0^\l{\rm d}x\,f(x)f(\l-x)F(\l-x)\approx\frac{3}{2}f(\l),
\ee
for $\l$ large.
Inserting this estimate into~(\ref{eq:Qndip}) leads to the announced result~(\ref{eq:Quni}).

The statistics of the number of records $M_n$ for subexponential underlying distributions $f(x)$
will be investigated at the end of section \ref{sec:psubex}.

\subsection{ Exponentially broad distributions}
\label{sec:eb}

We now consider the limiting class of exponentially broad distributions, defined by setting
\be
X=\e^{\Lambda T},
\ee
where $\Lambda$ is parametrically large, whereas $T$ has a fixed given distribution $g(t)$.
Exponentially broad distributions play a part
in the study of strongly disordered systems
(see \cite{CH,LD} and the references therein).
An explicit example is provided by the power-law distribution (see table~\ref{tab:distribs})
in the limit where the index $\theta$ goes to zero,
with the identification $\Lambda=1/\th$ and $g(t)=\e^{-t}$.
The overwhelming simplification brought by exponentially broad distributions
in the $\Lambda\to\infty$ limit is that $X_1<X_2$ is equivalent to $X_1\ll X_2$.
In other words, the distribution is so broad that,
if two independent variables $X_1$ and $X_2$ are drawn from the latter,
one is negligible with respect to the other with very high probability.

Considering exponentially broad distributions in the $\Lambda\to\infty$ limit
is useful for our purpose in several regards.
First, the exact probability of record breaking $Q_n$ can be derived in this limit, even for finite $n$.
Second, as we shall see, the derivation gives an insight on the clustering of records underlying
the non-trivial limit (\ref{eq:lim32}).
Third, this approach will be readily extended to higher values of the window width $p$ in section \ref{sec:p},
where other techniques are not available any more.

Within this setting, it is easy to derive the probability of record breaking $Q_n$.
We recall that $Q_n$ is the probability of having $Y_n>\max(Y_2,\dots,Y_{n-1})$,
with
\be
Y_n=X_{n-1}+X_n,\quad
Y_{n-1}=X_{n-2}+X_{n-1},
\ee
and so on.
If the variables $X$ are drawn from an exponentially broad distribution,
only two events contribute to $Q_n$:

\begin{enumerate}

\item[$\circ$]
The largest of the first $n$ $X$-variables is $X_n$.
This occurs with probability $1/n$.
In the $\Lambda\to\infty$ limit,
the variable $Y_n$ is also larger than all previous ones with certainty.
Hence the contribution $1/n$ to $Q_n$.

\item[$\circ$]
The largest of the first $n$ $X$-variables is $X_{n-1}$.
This again occurs with probability~$1/n$.
The condition $Y_n>\max(Y_2,\dots,Y_{n-1})$ reduces to $X_n>X_{n-2}$,
so that the relative probability of that event is $1/2$.
Hence the contribution $1/(2n)$ to $Q_n$.
\end{enumerate}
As a consequence, the probability of record breaking is exactly given by
\beq\label{eq:Qeb}
Q_n=\frac{3}{2n},
\eeq
for all $n\ge3$ and all exponentially broad distributions in the $\Lambda\to\infty$ limit.

The formula~(\ref{eq:Qeb}) gives both the exact value of $Q_n$ for exponentially broad distributions
and its asymptotic decay law (see~(\ref{eq:lim32}),~(\ref{eq:Quni}))
for all distributions with a subexponential tail.
The above derivation also demonstrates that the excess in the probability of record breaking (\ref{eq:Qeb})
with respect to the iid situation is due to a clustering of records.
The second event of the above list indeed yields two consecutive records.
Finally, the data shown in figure~\ref{fig:nqn} suggest that~(\ref{eq:Qeb})
provides an absolute upper bound for $Q_n$.
It is indeed quite plausible that the quantity $nQ_n$ plotted in figure~\ref{fig:nqn}
converges to the constant $3/2$ from below in the $\th\to0$ limit, uniformly in $n$.

\section{Extension to higher values of $p$}
\label{sec:p}

In this last section we consider sequences~(\ref{eq:Yp})
obtained by taking the moving average of a sequence of iid variables $X_i$
over an arbitrary finite window width $p\ge2$.
We shall mainly focus on the behavior of the probability of record breaking,
that we now denote by
\be
Q_n^\pe=\prob(Y_n>L_{n-1}).
\ee
The recursive structure of the problem described in section~\ref{sec:theo} still holds true,
however it becomes somewhat inefficient, as the number of variables is higher.
The recursion equation generalizing~(\ref{eq:recurF}) indeed involves a multiple integral over $p-1$ variables.
In particular, no exact solution is available any more.
In spite of this,
we shall be able to extend to higher values of $p$
most results of interest derived so far for $p=2$.

\subsection{Universal values of the probability of record breaking}
\label{sec:pfirst}

The first few values of $Q_n^\pe$ are universal,
i.e., independent of the underlying distribution $f(x)$.
Their values can be derived along the lines of reasoning of section~\ref{sec:firstn},
using again the Sparre Andersen theorem.
The first case of interest is $n=p$, where
\beq\label{qp0}
Q_p^\pe=1.
\eeq
There is indeed always a record at $n=p$,
as $Y_p$ is the first complete sum of $p$ terms.
For $n=p+1$, we have
\beq\label{qp1}
Q_{p+1}^\pe=\prob(X_{p+1}>X_1)=\frac{1}{2}.
\eeq
For $n=p+2$, we have
\beqa
Q_{p+2}^\pe&=&\prob(X_{p+2}>X_2,X_{p+1}+X_{p+2}>X_1+X_2)
\nonumber\\
&=&\prob(X_{p+2}-X_2>0,X_{p+2}-X_2+X_{p+1}-X_1>0)
\nonumber\\
&=&P_2=\frac{3}{8},
\label{qp2}
\eeqa
using the same argument as in section \ref{sec:firstn} for the derivation of $Q_4=P_2$ for $p=2$.
More generally, for $n=p+k$, with $1\le k\le p$, we have
\beq\label{qpuni}
Q_{p+k}^\pe=P_{k},
\eeq
where the expression of $P_k$ is given in~(\ref{eq:psa}).

The above formula
generalizes the results of section~\ref{sec:firstn} to an arbitrary window width $p\ge2$.
It exhausts the list of all universal values of the probability of record breaking.
In other words, $Q_{2p+1}^\pe$ is the first non-universal one, just as $Q_5$ for $p=2$.

\subsection{Superexponential and (some) exponential distributions}
\label{sec:pes}

The explanation given in section~\ref{sec:predict}
of the key dichotomy between~(\ref{eq:lim1}) and~(\ref{eq:lim32}),
based on the analytic structure of the Laplace transform $\hat f(s)$,
is not limited to $p=2$.
Its consequences are therefore expected to hold irrespective of the window width~$p$.

For superexponential distributions,
as well as for some exponentially decaying distributions,
we are therefore again led to compare the original problem to its cyclic variant
and to introduce a decoupled model,
where the random variables $Y_i$ of the original problem are now replaced by a sequence of iid random variables
\be
\Y_i=\stackunder{p\ \mathrm{ replicas}}{\underbrace{X_i+X'_i+X''_i+\cdots}},
\ee
generalizing (\ref{eq:Ycal}).
If the sum $\Y$ is atypically large,
then all its terms are atypically large as well, with very high probability.
We therefore predict a behavior of type~(\ref{eq:lim1}),~i.e.,
\be
nQ_n^\pe=1+\delta_n^\pe,
\ee
with the following estimate for the small relative correction $\delta_n^\pe$:
\beq\label{eq:delp}
\delta_n^\pe\approx n\int_0^\infty{\rm d}y\,\e^{-n\bar F_{p}(y)}
\int_0^y{\rm d}x\,f(x)f_{p-1}(y-x)\bar F(x),
\eeq
which is a direct generalization of~(\ref{eq:delheu}).
We again claim that this prediction is asymptotically correct, to leading order for large $n$,
whenever it decays to zero, i.e., essentially for all superexponential distributions.

The estimate~(\ref{eq:delp}) is now made explicit for a variety of distributions~$f(x)$.

\subsubsection{Pure exponential distribution.}

For an exponential distribution with density $f(x)=\e^{-x}$ and distribution function $F(x)=1-\e^{-x}$,
we have
\be
f_{p}(y)=\frac{y^{p-1}}{(p-1)!}\,\e^{-y}
\ee
as well as $\bar F_{p}(y)\approx f_{p}(y)$, to leading order for $y\gg1$,
and so~(\ref{eq:delp}) reads
\be
\delta_n^\pe\approx n\int_0^\infty{\rm d}y\,\frac{y^{p-2}}{(p-2)!}\,
\exp\left(-y-n\frac{y^{p-1}}{(p-1)!}\,\e^{-y}\right).
\ee
This integral can be evaluated along the lines of~(\ref{Qnint}) and~(\ref{eq:delnint}).
Omitting details, we obtain to leading order
\be
\delta_n^\pe\approx\frac{p-1}{\ln n}.
\ee
This estimate vanishes identically for $p=1$ and coincides with~(\ref{eq:dexpheu}) for $p=2$.
It demonstrates that the marginal character of the exponential distribution,
with its logarithmic correction term, persists to all higher values of $p$.

\subsubsection{Exponential distribution modulated by a power law.}
\label{sec:pexpui}

We now consider distri\-butions falling off as an exponential modulated by a power law, i.e.,
\beq
f(x)\approx\bar F(x)\approx A\,x^{a-1}\,\e^{-x}\quad(x\to\infty).
\label{pexpui}
\eeq

Along the lines of section \ref{sec:expui}, let us consider first the case where $a>0$.
We have
\be
f_p(y)\approx\bar F_p(y)\approx B_p\,y^{pa-1}\,\e^{-y}\quad(y\to\infty),
\ee
with $B_p=(A\Gamma(a))^p/\Gamma(pa)$.
Performing the integrals entering~(\ref{eq:delp}), we are left with the estimate
\beq
\delta_n\approx\frac{A\Gamma(pa)}{\Gamma((p-1)a)}\,\frac{1}{(\ln n)^a}.
\label{delp}
\eeq
This formula is a direct generalization of (\ref{delexpui}).
We thus conclude that exponential distributions of the form (\ref{pexpui}) with $a>0$
belong to the superexponential side of the dichotomy,
in the sense that (\ref{eq:lim1p}) holds,
with a correction falling off as~(\ref{delp}).
On the other hand, along the lines of section~\ref{sec:expui},
we are led to claim that exponential distributions with $a<0$ lead to (\ref{eq:lim32p}),
just as subexponential distributions.

\subsubsection{Distributions with bounded support and power-law singularity.}

In the case where $f(x)$ is supported by the interval [0, 1]
and has a power-law singularity of the form~(\ref{eq:power}) at its upper edge, we have
\be
\bar F_{p}(y)\approx B_p\eta^{pa},\quad f_{p}(y)\approx paB_p\eta^{pa-1},
\ee
with $\eta=p-y$ and $B_p=a^{p-1}(A\Gamma(a))^p/(p\Gamma(pa))$.
Performing the integrals enter\-ing~(\ref{eq:delp}),
we obtain a power-law decay for $\delta_n^\pe$, i.e.,
\beq
\delta_n^\pe\approx\frac{K(p,a)}{n^{1/p}},
\label{pwei}
\eeq
where the exponent only depends on the width $p$,
whereas the amplitude $K(p,a)$ reads
\be
K(p,a)=\frac{\Gamma(1/p)\,\Gamma(2a)\,\Gamma(pa)^{1+1/p}}{(pa)^{1-1/p}\,\Gamma(a)^2\,\Gamma((p+1)a)}.
\ee
This result extends~(\ref{eq:kofa}) to higher values of $p$.
The amplitude $K(p,a)$ has a local maximum for $a=0$, a local minimum for $a=1$,
and grows exponentially fast at large~$a$.
All these features hold irrespective of $p$, and survive in the formal $p\to\infty$ limit, i.e.,
\be
K(\infty,a)=\frac{\e^{-a}\,\Gamma(2a)}{a\,\Gamma(a)^2}.
\ee

\subsubsection{Distributions with exponential singularities.}

To close, we consider distributions with a bounded support
and an exponentially small singularity at their upper edge, of the form~(\ref{bse}),
as well as compressed distributions with
a superexponential tail extending up to infinity, of the form~(\ref{eq:compe}).

We again obtain a power-law decay for the correction $\delta_n^\pe$,
with continuously varying decay exponents $\omega_1(p,b)$ and $\omega_2(p,c)$,
which can be unified into a single monotonically decreasing function
\be
\omega(p,\alpha)=2\left(\frac{1+(p-1)2^{2\alpha/(1-\alpha)}}{p}\right)^{(\alpha-1)/(2\alpha)}-1
\ee
of the parameter $\alpha$ in the range $-1<\alpha<1$,
with the identifications~(\ref{bcalpha}).
We have in particular $\omega(p,-1)=1/p$, ensuring a smooth crossover with (\ref{pwei}),
$\omega(p,0)=\omega_1(p,\infty)=\omega_2(p,\infty)=2^{1/p}-1$
for the limiting situation of distributions with a double exponential fall-off,
and $\omega(p,1/3)=1/(2p-1)$, corresponding e.g.~to the half-Gaussian distribution.

\subsection{Exponentially broad distributions}
\label{sec:pse}

For exponentially broad distributions in the $\Lambda\to\infty$ limit,
the expression of $Q_n^\pe$ can be derived along the lines of section~\ref{sec:eb}.
We recall that $Q_n^\pe$ is the probability of having $Y_n>\max(Y_p,\dots,Y_{n-1})$,
with
\be
Y_n=X_{n-p+1}+\cdots+X_n,\quad
Y_{n-1}=X_{n-p}+\cdots+X_{n-1},
\ee
and so on.
If the $X$-variables are drawn from an exponentially broad distribution,
only the following events contribute to $Q_n^\pe$:

\begin{enumerate}

\item[$\circ$]
The largest of the first $n$ $X$-variables is $X_n$.
This occurs with prob\-ability $1/n$.
In the $\Lambda\to\infty$ limit,
the variable $Y_n$ is also larger than all previous ones with certainty.
Hence the contribution $1/n$ to $Q_n^\pe$.

\item[$\circ$]
The largest of the first $n$ $X$-variables is $X_{n-1}$.
This again occurs with prob\-ability~$1/n$.
The condition $Y_n>\max(Y_2,\dots,Y_n)$ reduces to $X_n>X_{n-2}$,
so that the relative probability of that event is $1/2$.
Hence the contribution $1/(2n)$ to~$Q_n^\pe$.

\item[$\circ$]
The largest of the first $n$ $X$-variables is $X_{n-2}$.
This again occurs with prob\-ability~$1/n$.
The condition $Y_n>\max(Y_2,\dots,Y_n)$ reduces to
\bea
X_n-X_{n-3}>0,
\\
X_n-X_{n-3}+X_{n-1}-X_{n-4}>0.
\eea
so that the relative probability of that event is $P_2=3/8$,
again by virtue of the Sparre Andersen theorem.
Hence the contribution $P_2/n$ to $Q_n^\pe$,
and so on.

\end{enumerate}

Summing up the probabilities of the events listed above,
we predict that the probability of record breaking is exactly given by
\beq\label{eq:Qebp}
Q_n^\pe=\frac{R_p}{n},
\eeq
for all $p\ge2$ and $n\ge2p-1$ and all exponentially broad distributions in the $\Lambda\to\infty$ limit.
The numerator of the above formula reads
\be
R_p=\sum_{k=0}^{p-1}P_k,
\ee
where the integer $k$ numbers the items of the above list
and where the expression of $P_k$ is given in~(\ref{eq:psa}).
Equation~(\ref{eq:piz}) yields
\be
\w R(z)=\sum_{p\ge1}z^p R_p=\frac{z}{1-z}\,\w P(z)=\frac{z}{(1-z)^{3/2}}.
\ee
The $R_p$ are therefore universal rational numbers given by
\beqa
R_p&=&\frac{(2p-1)!}{2^{2p-2}(p-1)!^2}=2pP_p=(2p-1)P_{p-1}
\nonumber\\
&=&1,\ \frac{3}{2},\ \frac{15}{8},\ \frac{35}{16},\ \frac{315}{128},\ \frac{693}{256},\dots,
\label{rpres}
\eeqa
for $p=1,2,3,4,5,6,\dots$, and growing as
\be
R_p\approx2\sqrt\frac{p}{\pi}
\ee
at large $p$.

The formulas~(\ref{qpuni}) and~(\ref{eq:Qebp}) overlap for two values of $n$,
namely $2p-1$ and~$2p$, for which they consistently predict
\be
Q_{2p-1}^\pe=P_{p-1}=\frac{R_p}{2p-1},\quad
Q_{2p}^\pe=P_p=\frac{R_p}{2p}.
\ee

\subsection{Subexponential distributions}
\label{sec:psubex}

Following the line of thought sketched in the very beginning of section~\ref{sec:pes},
we are led to extend the dichotomy between~(\ref{eq:lim1}) and~(\ref{eq:lim32}) to higher values of $p$,
and to predict the following asymptotic decay of the probability of record breaking at large~$n$:
\beq
Q_n^\pe\approx\frac{R_p}{n},
\label{qnpasy}
\eeq
for all $p\ge2$ and all subexponential distributions $f(x)$,
where the amplitude $R_p$ is predicted by the exact
analysis of the limiting case of exponentially broad distributions
(see section \ref{sec:pse}).
The latter amplitude, given by~(\ref{rpres}), is therefore universal,
in the sense that it only depends on the window width $p$.

The formula~(\ref{eq:Qebp}) therefore has the same status as~(\ref{eq:Qeb}).
It gives the exact value of $Q_n^\pe$ for exponentially broad distributions
in the $\Lambda\to\infty$ limit for all $n\ge2p-1$.
It is also expected
to describe the asymptotic decay law of $Q_n^\pe$ for all subexponential distributions,
and furthermore to provide an absolute upper bound for $Q_n^\pe$ for all $n\ge p$.

\section{Discussion}
\label{sec:disc}

This paper was devoted to the statistics of records for the moving average of a sequence of iid variables.
Most results concern the case where the window width is $p=2$.
The main emphasis has been put on the probability of record breaking $Q_n$ at time $n$,
and on the distribution of the number of records~$M_n$ up to time $n$.
In sections~\ref{sec:exp} to~\ref{sec:th1} we have given full analytical solutions of the problem
for three particular parent distributions: exponential, uniform and power-law with $\th=1$.
The exact results obtained there provide useful checks of the heuristic approach used
in the asymptotic analysis of the general situation (section~\ref{sec:gal})
and in its extension to higher values of $p$ (section~\ref{sec:p}).

Quite serendipitously,
the three distributions which have lent themselves to an exact analytical treatment
are prototypical in several regards.
First, each of them is a representative of one of the three universality classes of extreme value statistics:
Weibull, Gumbel and Fr\'echet.
Second, they are also representatives of the dichotomy, as regards the properties of records for the moving average, between
superexponential distributions, where the product $nQ_n$ tends to unity
and the distribution of the number of records is asymptotically Poissonian,
and subexponential distributions,
where $nQ_n$ admits the non-trivial universal limit 3/2, or more generally $R_p$,
and the distribution of the number of records exhibits novel universal clustering features.
The uniform and power-law distributions are respectively typical of the superexponential and subexponential classes,
whereas the exponential distribution is a representative of the exponential class,
which is marginal and split on both sides of the dichotomy, as seen in section~\ref{sec:expui}.

\begin{figure}[!ht]
\begin{center}
\includegraphics[angle=0,width=.85\linewidth]{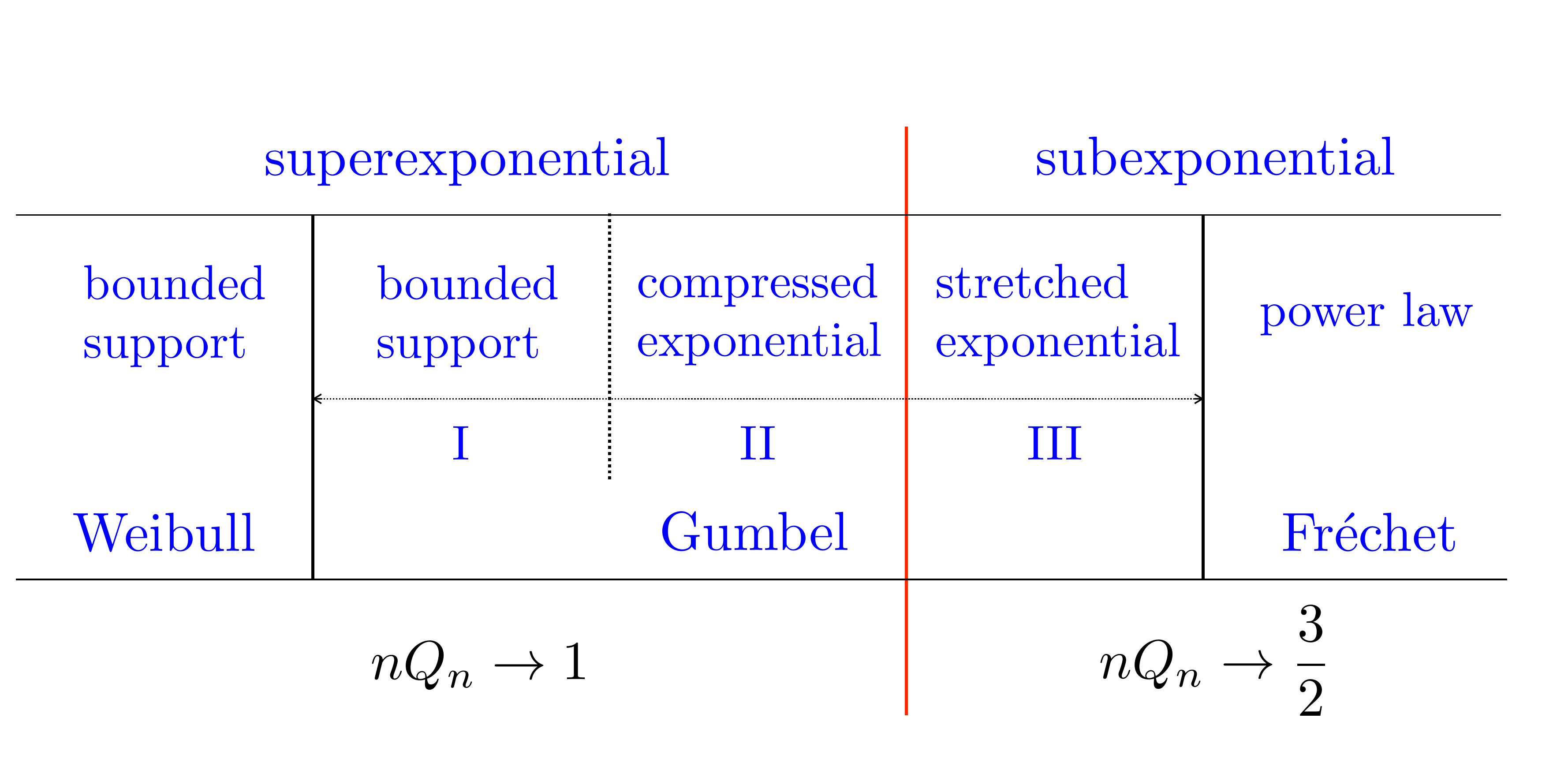}
\caption
{Sketchy representation of the realm of parent probability distributions (see~text for details).
The last line summarizes the results for the case $p=2$.}
\label{fig:diag}
\end{center}
\end{figure}

Our main results can be summarized
in the sketchy representation of the realm of parent probability distributions shown in figure~\ref{fig:diag}.
The tail of the distribution is more and more heavy,
i.e., the density $f(x)$ falls off more and more slowly,
as one progresses from left to right.
The red line in figure~\ref{fig:diag} represents the boundary of the dichotomy,
with superexponential distributions to its left and subexponential distributions to its right,
with the marginal class of exponential distributions sitting on the line itself.
To the left of the red line,
the product $nQ_n$ tends to unity, just as for records of iid variables.
Superexponential distributions can be classified according to the exponent~$\omega$
describing the power-law decay $\delta_n\sim n^{-\omega}$
of the correction such that $nQ_n=1+\delta_n$.
For distributions in the Weibull class,
i.e., with a bounded support and a power-law singularity at its upper end,
$\omega$ is constant and equal to 1/2, and more generally $1/p$.
For superexponential distributions in the Gumbel class,
whose support is either bounded (Region~I) or unbounded (Region~II),
the exponent $\omega(b)$ or $\omega(c)$ decreases from 1/2 to~0,
and more generally $\omega(p,b)$ or $\omega(p,c)$ decreases from $1/p$ to 0.
To the right of the red line,
the product $nQ_n$ admits the non-trivial universal limit 3/2, and more generally $R_p$.
This limit holds both for subexponential distributions falling off faster than any power of $x$
(Region~III of the Gumbel class)
and for those exhibiting a power-law tail (Fr\'echet class).

The key dichotomy highlighted in the present work for the properties of records of the moving average,
between~(\ref{eq:lim1}) and~(\ref{eq:lim32}) (or more generally between~(\ref{eq:lim1p}) and~(\ref{eq:lim32p})),
i.e., essentially
between the subexponential and superexponential classes of distributions, appears as very robust.
It is therefore expected to have far-reaching consequences on other quantities,
besides the probability of record breaking $Q_n$ and the number of records $M_n$.
Consider the example of the distribution of the maximum $L_n$ of the first $n$ daughter $Y$-variables.
The heuristic approach put forward in section~\ref{sec:gal}
suggests that the distribution of $L_n$ is close to that of the maximum of $n$ iid $X$-variables
for subexponential distributions, to the right of the red line,
whereas it is close to that of the maximum of $n$ iid variables of the form $Y=X+X'$
for superexponential distributions, to the left of the red line.
This claim is corroborated by the exact or asymptotic expressions
for the mean or median values of $L_n$ derived in sections~\ref{sec:exp} to~\ref{sec:th1}.

Let us close with a word on more general linear filters of the form
\be
Y_n=\sum_{k\ge0}K_kX_{n-k},
\ee
used e.g.~in digital signal processing,
transforming a sequence of iid random variables~$X_n$ to a filtered sequence $Y_n$,
whose entries are clearly not iid any more.
Many open questions of interest related to extremes and records in such filtered sequences could be addressed.
It can be anticipated that the occurrences of records will exhibit some clustering,
especially if the distribution $f(x)$ of the parent variables is broad enough,
even though a clear-cut universal dichotomy is not to be expected in general.

\ack
CG is grateful to S Majumdar for arousing his interest in the statistics of records for sequences made of sums of successive iid random variables.

\end{document}